\documentclass[letterpaper,twocolumn,10pt]{article}
\usepackage{usenix-2020-09}

\usepackage{tikz}
\usepackage{amsmath}

\usepackage{filecontents}

\usepackage{tikz}
\usepackage{amsmath}
\usepackage{colortbl}
\usepackage{tikz}
\usepackage{marvosym}

\usepackage{subcaption}
\usepackage{booktabs} 

\usepackage{threeparttable}
\usepackage[available]{usenixbadges}
\usepackage{graphicx}
\usepackage[many]{tcolorbox}
\usepackage{mathtools,latexsym,amsfonts,stmaryrd}
\usepackage{pbox}
\usepackage{amsthm}
\usepackage{multirow}
\usepackage{tikz}
\usepackage{mathrsfs}
\usepackage{mathpartir}
\usepackage[ruled,linesnumbered,vlined,noend]{algorithm2e}
\usepackage{url}
\usepackage{syntax}
\usepackage{framed}
\usepackage[noend]{algpseudocode}
\usepackage{flushend}
\usepackage{listings}
\usepackage{moresize}
\usepackage{wrapfig}
 
\usepackage{seqsplit}
\usepackage{alltt}
\usepackage{xspace}
\usepackage[normalem]{ulem}
\usepackage{enumitem}
\usepackage{pifont}
\DeclareMathAlphabet{\mathcal}{OMS}{cmsy}{m}{n}

\usepackage{amsmath, listings, amsthm, proof, xspace}

\usepackage{thmtools}

\declaretheoremstyle[spaceabove=\topsep,notefont=\normalfont\itshape]{mystyle}

\definecolor{ForestGreen}{RGB}{34,139,34}

\newcommand{\revise}[2]{{\color{red}{\ifx&#1&\else- #1\fi}} {\color{ForestGreen}{\ifx&#2&\else+ #2\fi}}}%
\renewcommand{\revise}[2]{#2}%

\usetikzlibrary{arrows,patterns, decorations.pathreplacing}

\newcommand{\F}{Fig.}

\newcommand{\T}{Table}
\renewcommand{\S}{Sec.}
\newcommand{\A}{Alg.}

\newcommand{\ignore}[1]{}
\newcommand{\mysubref}[2]{\hyperref[#1]{\ref*{#1}(#2)}}

\lstdefinestyle{base}{
  moredelim=**[is][\color{red}]{@}{@},
  escapeinside={<@}{@>}
}

\usepackage{pifont}

\newcommand\DejaVuttfamily{%
  \fontfamily{DejaVuSansMono-TLF}\selectfont
}

\lstdefinestyle{base}{
  moredelim=**[is][\color{red}]{@}{@},
  escapeinside={<@}{@>}
}

\lstdefinelanguage
   [x64]{Assembler}     
   [x86masm]{Assembler} 
   {morekeywords={CDQE,CQO,CMPSQ,CMPXCHG16B,JRCXZ,LODSQ,MOVSXD, %
                  POPFQ,PUSHFQ,SCASQ,STOSQ,IRETQ,RDTSCP,SWAPGS, %
                  rax,rdx,rcx,rbx,rsi,rdi,rsp,rbp, %
                  r8,r8d,r8w,r8b,r9,r9d,r9w,r9b,reg128,m128}} 

\let\OLDthebibliography\thebibliography
\renewcommand\thebibliography[1]{
  \OLDthebibliography{#1}
  \setlength{\parskip}{0pt}
  \setlength{\itemsep}{1pt plus 0.85ex}
}
\usepackage{listings}
\usepackage{fancyvrb}
\usepackage{color}
\definecolor{lightgray}{rgb}{.9,.9,.9}
\definecolor{darkgray}{rgb}{.4,.4,.4}
\definecolor{purple}{rgb}{0.65, 0.12, 0.82}
\definecolor{commentgreen}{RGB}{63,127,95}
\definecolor{pyblue}{RGB}{59,117,175}
\definecolor{pyorange}{RGB}{239,134,54}
\definecolor{pygreen}{RGB}{81,158,62}

\usepackage{xcolor}
\colorlet{myPurple}{blue!40!red}
\definecolor{myOrange}{RGB}{255,192,0}

\lstdefinelanguage{Solidity}{
  keywords={len,delete,int,void,payable, public, event, contract, typeof, new, true, false, catch, function, return, null, catch, switch, var, if, while, do, else, case, break,struct,const,socklen_t,sa_familty_t,char,sockaddr,load},
  keywordstyle=\color{violet}\bfseries,
  ndkeywords={class, export, boolean, throw, implements, import, this},
  ndkeywordstyle=\color{darkgray}\bfseries,
  identifierstyle=\color{black},
  sensitive=false,
  comment=[l]{//},
  escapeinside={(*@}{@*)},          
  morecomment=[s]{/*}{*/},
  commentstyle=\color{commentgreen}\ttfamily,
  stringstyle=\color{red}\ttfamily,
  morestring=[b]',
  morestring=[b]"
}

\newcommand{\rnum}[1]{\uppercase\expandafter{\romannumeral #1\relax}}

\usepackage{xcolor}

\algnewcommand{\LeftComment}[1]{\Statex \(\triangleright\) #1}

\definecolor{pptbrown}{RGB}{132,60,12}
\definecolor{pptgreen}{RGB}{56,87,35}
\definecolor{pptred}{RGB}{155,30,20}
\definecolor{pptdy}{RGB}{127,96,0}

\newcommand{\parh}[1]{\noindent\textbf{#1}}

\newcommand{\tool}{\textsc{G$^2$Fuzz}\xspace}
\newcommand{\name}{\tool}

\newcommand{\rom}[1]{\uppercase\expandafter{\romannumeral #1\relax}}

\begin{document}

\date{}

\title{Enabling Low-Cost and Comprehensive Grammar-Aware Fuzzing with Large Language Models\footnote{}}
\title{\name: Enabling Low-Cost and Comprehensive Non-Textual Input Fuzzing with LLM-Synthesized Input Generators}
\title{Low-Cost and Comprehensive Non-textual Input Fuzzing with LLM-Synthesized Input Generators\thanks{This paper has been accepted by USENIX Security 2025.}}

\author{
{\rm Kunpeng Zhang\textsuperscript{1}} \quad 
{\rm Zongjie Li\textsuperscript{1}} \quad 
{\rm Daoyuan Wu\textsuperscript{1\textsuperscript{\Letter}}} \quad 
{\rm Shuai Wang\textsuperscript{1\textsuperscript{\Letter}}} \quad 
{\rm Xin Xia\textsuperscript{2}} \quad 
\and
\textsuperscript{1}The Hong Kong University of Science and Technology \\
\textsuperscript{2}Zhejiang University \\
}

\maketitle

\begin{abstract}
Modern software often accepts inputs with highly complex grammars. To
    conduct greybox fuzzing and uncover security bugs in such software, it is
    essential to generate inputs that conform to the software input grammar.
    However, this is a well-known challenging task because it requires a deep
    understanding of the grammar, which is often not available and hard to
    infer. Recent advances in large language models (LLMs) have shown that they
    can be used to synthesize high-quality natural language text and code that
    conforms to the grammar of a given input format. Nevertheless, LLMs are
    often incapable or too costly to generate \textit{non-textual outputs}, such
    as images, videos, and PDF files. This limitation hinders the application of
    LLMs in grammar-aware fuzzing.
    
    We present a novel approach to enabling grammar-aware fuzzing over
    non-textual inputs. We employ LLMs to synthesize and also mutate
    \textit{input generators}, in the form of Python scripts, that
    generate data conforming to the grammar of a given input format. Then,
    non-textual data yielded by the input generators are further mutated by
    traditional fuzzers (AFL++) to explore the software input space effectively.
    Our approach, namely \tool, features a hybrid strategy that combines a
    ``holistic search'' driven by LLMs and a ``local search'' driven by
    industrial quality fuzzers. Two key advantages are: (1) LLMs are
    good at synthesizing and mutating input generators and enabling jumping out
    of local optima, thus achieving a synergistic effect when combined with
    mutation-based fuzzers; (2) LLMs are less frequently invoked unless really
    needed, thus significantly reducing the cost of LLM usage.
    We have evaluated \tool on a variety of input formats, including TIFF
    images, MP4 audios, and PDF files. The results show that \tool
    outperforms SOTA tools such as AFL++, Fuzztruction, and FormatFuzzer in
    terms of code coverage and bug finding across most programs tested on three
    platforms: UNIFUZZ, FuzzBench, and MAGMA. \tool also discovers 10 unique
    bugs in the latest real-world software, of which 3 are confirmed by CVE.\let\thefootnote\relax\footnote{{\textsuperscript{\Letter}~}Corresponding authors.}
\end{abstract}

\section{Introduction}
\label{sec:introduction}



%

Modern software often accepts inputs with highly complex grammars, such as
images, configuration files, and network packets. Fuzzing such software is
well-known to be challenging~\cite{godefroid2008grammar, sargsyan2018grammar,
eberlein2020evolutionary, hodovan2018grammarinator,le2021saffron, liu2023vd} because it
requires a deep understanding of software grammar to fully explore the input space. Often, one needs to prepare sample
inputs that conform to the grammar of the input format and also exhibit a
variety of characteristics before conducting effective greybox
fuzzing~\cite{lyu2019mopt, lyu2022ems, fioraldi2020weizz, yue2020ecofuzz,
lemieux2018fairfuzz, aschermann2019redqueen, she2024fox} and uncover security bugs.

Recent structure-aware fuzzers have explored solutions to alleviate the
above challenges with inference-based fuzzing and grammar-aware fuzzing, yet
they still have limitations. 
Inference-based fuzzers, such as ProFuzzer~\cite{profuzzer},
FuzzInMem~\cite{liu2024fuzzinmem}, and WEIZZ~\cite{fioraldi2020weizz}, can infer
input grammars and generate inputs on-the-fly. However, they often suffer from
low accuracy and weak scalability, by inferring simple input fields and
struggling to mutate file structures. ProFuzzer and WEIZZ are time-consuming for
long inputs, while FuzzInMem requires programs having printer functions that
convert in-memory data structures to files.
Grammar-aware greybox fuzzers~\cite{guo2013gramfuzz, park2020fuzzing, superion,
aschermann2019nautilus, blazytko2019grimoire} often require pre-knowledge of the
input grammar (e.g., provided by the users). Such information is often not
available or incomplete, making it obscure to comprehensively understand input
fields and their relations. Moreover, current fuzzing approaches primarily
validate basic structural fields like size and checksums in file formats but
neglect complex features. These features can trigger more complex logic,
revealing deeper bugs. Complex features often include intricate chunks or
constraints between chunks, posing challenges to traditional fuzzing methods.
Fuzztruction~\cite{bars2023fuzztruction}
mitigates this challenge from a different perspective by injecting
faults into generator applications to produce inputs with highly
complex formats. However, Fuzztruction relies on the availability of a suitable
generator application, which still requires experienced researchers to
identify.

Large language models (LLMs) are transformer-based neural networks that have
achieved state-of-the-art (SOTA) performance in natural language and code processing tasks. Thus, one may expect that LLMs could generate input samples with various valid grammars, thus driving grammar-aware
fuzzing on its own. Indeed, we have seen some recent works that leverage LLMs to
generate inputs for fuzzing~\cite{xia2024fuzz4all, deng2023large, deng2024large, meng2024large}. Nevertheless, we clarify that although LLMs
are capable of generating \textit{textual} outputs, such as natural language
text and code, we find that LLMs are often incapable or too costly of generating
\textit{non-textual} data samples as required by many software. We present a
detailed analysis in \S~\ref{sec:motivation}.

Instead of instructing LLMs to directly generate non-textual fuzzing inputs,
this paper explores another perspective to augment mutation-based fuzzing with
LLMs. The key idea is to leverage LLMs to automatically synthesize and
further mutate input generators (often in the form of Python scripts)
customized to the specific features and structures of the target file format. By
executing these generators, we can produce inputs that exhibit a wide range of
features and structures, potentially triggering different program logic and
exploring previously untested code regions. Moreover, these generated
non-textual inputs can be rapidly mutated by traditional mutation-based fuzzers,
such as AFL++, to effectively explore the input space. Holistically, this
approach offers a new and unique hybrid view to augment fuzzing with LLMs:
\textit{LLMs are particularly good at synthesizing distinct input generators and
enabling the escape from ``local optima,'' whereas mutation-based fuzzers excel
at conducting fine-grained, local searches in the input space efficiently.} We
show that our novel combination of LLMs and mutation-based fuzzers can achieve a
synergistic effect, leading to a significant improvement in code coverage and
bug finding. Moreover, since we only invoke LLMs when necessary to synthesize
new input generators, we substantially reduce the cost of LLM usage.

We implement the above approach in a novel fuzzing framework, namely
\tool.\footnote{\tool\ stands for ``\textbf{g}rammar-aware fuzzing with
LLM-synthesized input \textbf{g}enerators''.} When users specify an input format
name (e.g., ``TIFF''), \tool\ employs de facto LLMs, such as
GPT-3.5 and llama-3-70b-instruct, to automatically synthesize input generators in Python
scripts that generate TIFF images. \tool\ facilitates several strategies to
further mutate the synthesized generators.
Then, \tool\ executes generators to produce a diverse set of non-textual inputs,
and also employs AFL++ to mutate the synthesized inputs. When the employed
fuzzers fail to uncover new code coverage to a certain extent, \tool\ invokes
LLMs to synthesize new and distinct input generators, and then further mutates
the generated non-textual inputs using AFL++. This process continues until the
target software is fully covered or a certain time budget is reached.

We evaluate \tool\ on 34 input formats,
including JPEG images, TIFF images, and MP4 videos. Our results
show that \tool\ can consistently outperform SOTA mutation-based
fuzzers, such as AFL++, and several structure-aware fuzzer baselines, in terms
of code coverage and bug finding. We evaluated it on three third-party
benchmarks: UNIFUZZ~\cite{unifuzz}, FuzzBench~\cite{fuzzbench}, and MAGMA~\cite{magma}. Our results show that \tool\ achieves
the best performance in code coverage and bug finding across all three
platforms.
We find that with the help of LLMs, \tool is able to discover many unique edge/function coverage that other fuzzers cannot find.
Moreover, we show that \tool\ incurs a very low cost of LLM usage;
fuzzing a target software with GPT-3.5 for 24 hours only costs less than 0.2\$ in LLM usage.
We have used \tool\ to find 10 unique bugs in the latest real-world software, of which 3 are confirmed by CVE.
%
In sum, our contributions are as follows: 

\begin{itemize}[leftmargin=*,noitemsep,topsep=0pt]
    \item We introduce a novel approach to augmenting mutation-based  
        fuzzing using LLMs. The core idea is to combine the strengths of LLMs in
        synthesizing and mutating diverse input generators and the strengths of
        mutation-based fuzzers in performing fine-grained mutations over
        non-textual data. This approach leverages a synergistic effect to
        deliver effective fuzzing at a moderate cost.

    \item We design \tool\ that concretizes the above idea. \tool\ properly and
        periodically invokes LLMs and mutation-based fuzzers to benefit from
        their respective strengths. \tool\ features a set of design principles
        and optimizations to make it highly efficient and practical.

    \item Our results show that \tool\ consistently outperforms SOTA
        mutation-based fuzzers and several other fuzzer baselines in terms of
        code coverage and bug finding across various input formats and testing 
        platforms. \tool\ has discovered 10 unique bugs in the latest
        real-world software.
\end{itemize}

\section{Preliminaries}
\label{sec:preliminary}

\parh{Large Language Models (LLMs).}~LLMs, transformer-based neural networks,
have reached SOTA performance in various NLP tasks, including translation and
summarization. Autoregressive (e.g., GPT) and masked language modeling (e.g.,
BERT) are essential for textual output, while models like CLIP\cite{li2022clip}
and DALL-E\cite{ramesh2021zero} handle non-textual data, enhancing their range
of applications. The community has noted that LLMs have the potential to augment
software fuzzing~\cite{Brendan}.

\parh{Greybox Fuzzing.}~Greybox fuzzing, a technique for finding software security bugs, relies on lightweight instrumentation for execution feedback to mutate inputs more effectively. 
Fuzzers like AFL\cite{afl}, AFL++\cite{afl++}, and Honggfuzz~\cite{honggfuzz} have advanced this field. AFL++, with optimizations such as Redqueen, is recognized as the de facto standard fuzzer, widely used by the security community to detect bugs.


\parh{Grammar-Aware Fuzzing.}
Grammar-aware fuzzing, a form of greybox fuzzing, produces inputs based on
precise grammar rules, effectively identifying vulnerabilities in software that
handle complex input structures.
Tools like FormatFuzzer~\cite{formatfuzzer}, Gramatron~\cite{gramatron}, and
Superion~\cite{superion} leverage provided grammars to uncover security bugs in
real-world software. To generate inputs in highly complex formats,
Fuzztruction~\cite{bars2023fuzztruction} deliberately injects faults
into generator applications.



\parh{Inference-Based Fuzzing.}~Inference-based fuzzing, such as
ProFuzzer~\cite{profuzzer}, GreyOne~\cite{greyone}, and WEIZZ~\cite{fioraldi2020weizz},
leverages inferred relationships between input bytes and path constraints to
generate targeted test inputs.
This method analyzes internal logic and data formats to create relevant test
cases, enhancing coverage and reducing noise in results. 

\section{Motivation}
\label{sec:motivation}

\subsection{Related Work and Limitations}

Existing methods can be categorized into two types based on the input they
handle: (1) Text-format fuzzing, and (2) Binary-format fuzzing. Text-format
fuzzing primarily tests programs using text inputs, such as
Superion~\cite{superion}, Nautilus~\cite{aschermann2019nautilus}, and
Grimoire~\cite{blazytko2019grimoire}. These methods generate a variety of valid
text inputs based on provided specifications, including formats like XML,
 Ruby, SQL, and SMT. On the other hand, binary-format fuzzing
tests programs with binary inputs, such as FormatFuzzer~\cite{formatfuzzer},
FuzzInMem~\cite{liu2024fuzzinmem}, WEIZZ~\cite{fioraldi2020weizz}, and
AFLSmart~\cite{pham2019smart}. These methods split the input into multiple
chunks and perform mutations on these chunks to create diverse inputs, such as
JPEG, PDF, TIFF, MP3, and MP4.

\tool belongs to the latter category, constructing binary format files with
complex features for exploring deeper into the code. Despite the significant
advancements made by existing methods, they still face the following three
challenges:

\noindent \textbf{Challenge I: Generating Files with Complex Features.} The
current approaches focus primarily on the basic structure of target file
formats, such as generating valid basic structural fields like size fields,
checksums, and bitfields. However, a target binary file format often
incorporates various complex features, and per our observation (see 
\S~\ref{sec:unifuzz}), files with complex features often have the potential to
trigger more complex program logic, thereby likely uncovering deeper-seated
bugs. Compared to the basic structures, these complex features differ mainly in
two aspects: 1) complex features likely introduce extra complex chunks in the
binary file, and 2) varying constraints (e.g., numerical constraints raised by
checksum) may be introduced among binary file chunks. Additionally, certain
dependencies exist among different (basic/complex) features, where one feature
depends on another to be implemented. For instance, to enable JPEG compression
in a TIFF file, the file must first support the ``YCbCr/RGB color space''
feature. All these scenarios pose major challenges for existing binary-format
fuzzers. For example, current fuzzers fail to generate TIFF files with LZW data due to inaccurate inference (e.g., WEIZZ) or incomplete grammars lacking LZW syntax (e.g., FormatFuzzer, AFLSmart), limiting their parsing and mutation capabilities. See Appendix~\ref{sec:caseStudy} for more details.

\noindent \textbf{Challenge II: Require Format Specifications and Manual
Coding.}~Previous works often rely on provided format specifications or human
effort to modify code, as seen in FormatFuzzer~\cite{formatfuzzer}. FormatFuzzer
obtains format templates from the 010 Editor repository and uses them for
parsing. However, manual coding is still required for the generation process.
Furthermore, modifying complex formats like MP4 can take ``over a week''
(per~\cite{formatfuzzer}), due to its multiple chunk types, many of which are
not fully detailed in the original binary template.
Additionally, Fuzztruction can generate diverse files by injecting faults into generator applications. Yet, it relies
on experienced researchers to manually identify and instrument suitable
generator applications, and finding appropriate generators for less common
formats is often challenging. Here, we search GitHub for generator applications
for all 34 formats listed in~\T~\ref{tab:formats}. The search uses the keywords
\textit{"FORMAT converter/transformer/generator language:C++ stars:>5"}, where
 ``"FORMAT'' serves a placeholder for specific formats. For example, for JPG files,
one of the search queries is \textit{"JPG converter language:C++ stars:>5"}.
Generators were found for 21 formats (usability untested), while no generators
were available for the remaining 13 formats.

\noindent \textbf{Challenge III: Simultaneously Process Multiple Formats.} Many
software can process multiple input formats. However, existing grammar-aware
fuzzers typically generate files of a single format during the fuzzing process.
This limitation hampers their effectiveness in thoroughly testing software that
accept diverse input formats, potentially missing bugs related to the handling
of specific file types.

One solution may be launching multiple fuzzers in parallel, each focusing on one
input format, and then aggregating the individual fuzzing results at the end.
However, programs often include routine code for preprocessing and error
handling that are independent of specific file formats. Parallelism can result
in repetitive efforts in these common routines, not only consuming time but also
wasting resources.


\noindent \textbf{Insight.} We view that the aforementioned limitations can be
addressed by cleverly leveraging LLMs. Our insights are as follows: for
generating files with complex features (\textbf{Challenge I}), numerous
libraries for file generation are already available online. 
These libraries offer APIs to directly construct complex features of the target format.
Since LLMs have been trained on vast datasets that
presumably include these online codebases, they shall be able to yield binary
file generation scripts (code in Python) tailored to the required file features.
By running this generator, we can produce files that exhibit the desired
features. For example, to implement LZW compression for TIFF
(\F~\ref{fig:seedCompress2}), we can employ LLMs to construct the corresponding
generator with 3 lines of code, as in \F~\ref{fig:tiffCompressCode}. With LLMs,
common complex file structures can be generated with a moderate amount of code.

Using LLMs to generate generators evidently eliminates human efforts or the need
for preparing format specifications. This enables a fully automated testing
process, whereas existing methods require manual coding and format preparation
(\textbf{Challenge II}). Moreover, our fuzzing pipeline maintains generators of
different binary formats unifiedly (\S~\ref{sec:synthesis}), allowing
simultaneously processing multiple formats yet largely reducing repetitive
efforts (\textbf{Challenge III}). This allows us to concentrate resources and
efforts on more in-depth testing of code areas that are closely related to
different file formats.

\subsection{The Pilot Study of LLMs}

LLMs have been extensively trained using large-scale datasets, enabling them to
learn complex patterns and generate high-quality outputs. This extensive
pre-training enables LLMs to excel in various open-ended, structured data
generation tasks, such as code completion and generation~\cite{xu2022systematic, liu2020multi, guo2023longcoder, dinh2024large, dinh2024large, pilault2020extractive, zhang2024benchmarking, tang2023evaluating, pearce2023examining, ahmad2024hardware}, text to image
translation~\cite{lu2024llmscore, gani2023llm, huang2023towards, qu2024unified, alabdulrahman2024sarid}, and QA tasks for customer support~\cite{rangapur2024battle, kim2024learning, saito2024unsupervised}. It is believed that the vast
amount of training data helps LLMs capture the nuances of language and produce
accurate and contextually appropriate results. 

\noindent \textbf{Limitation of De Facto LLMs.}~We are positive that LLMs can be
used to augment mutation-based fuzzing, given that LLMs may possess complex
grammatical knowledge to facilitate continuous testing of software with complex
input formats. However, we find that LLMs are often less capable or even
incapable of generating \textit{non-textual} outputs. In particular, while
modern fuzz testing frequently targets non-textual inputs like image processing
libraries, general-purpose LLMs are not designed to generate such non-textual
data. Moreover, while cutting-edge LLMs such as DALL-E~\cite{dalle} can generate
images, our tentative exploration shows that the image format is limited to a
small set of predefined formats. DALL-E only supports generating images in
common image formats, such as PNG and GIF, even if we specify requiring other
formats (TIFF, RAW, BMP, etc.) in the prompts. Moreover, DALL-E usage costs
\$40.00 per 1,000 images, making it expensive for large-scale fuzzing.
Generating an input sample can take several seconds, significantly impacting
fuzzing throughput. Besides images, other non-textual file types, such as MP4
for video, MP3 for audio, PDF for documents, and Binary Large Object (BLOB), are
often not supported by existing LLMs. Finding all customized LLMs can be
challenging. 

We analyze the input format distributions of FuzzBench programs.
FuzzBench~\cite{fuzzbench} is one most widely-used benchmarking platform
developed by Google to evaluate fuzzing. It includes many widely used
open-source projects that process a variety of input formats. We believe the
analysis results will be generalizable due to the size and diversity of the
benchmarks. 
We find that \textit{73\% of the programs only accept non-textual inputs.}
Programs that accept non-textual inputs are more common than those accepting
textual inputs in traditional fuzzing. For these programs, general LLMs cannot
directly generate fuzzing inputs (reasons discussed before). Moreover, while
some cutting-edge LLMs can generate PNG and JPEG inputs, other formats still
lack support.

\noindent \textbf{LLM-Enabled Opportunity.}~We observe that most binary files
can be produced using Python libraries. For example, we can use \textsc{PIL} to
generate JPEG files with different structures and \textsc{cv2} to generate PNG
files. Since documents for these libraries are likely included in the LLM
training data, it is reasonable to expect that LLMs can synthesize generators
for binary files based on these libraries. 
In this step, we conducted experiments on the available formats in UNIFUZZ, FuzzBench, and MAGMA to test whether LLMs can generate a generator for each format. For more details, refer to~\S~\ref{sec:HandledFormats}. 
In short, all these non-textual data can be generated by Python scripts.
Consequently, we can use LLMs to synthesize various generators, producing
different structured inputs and exploring deeper code regions. However, we have
to address the following two challenges.

\parh{Technical Challenge I: Diversity.}~We find that LLM outputs are
often less diverse and uneasy to control; this is undesirable in fuzzing, which
expects a large number of generators that are diverse and cover as much 
of the input space as possible. 
Overall, our tentative exploration shows that LLM outputs are often predictable,
meaning that the software under fuzzing may process many similar inputs that do
not effectively cover the input space. 

During our preliminary study, we attempted to calibrate the diversity of
synthesized generators with several tactics, such as temperature control and
top-$k$ sampling, but the results were not satisfying. For example, while one
may instruct LLMs to ``\textit{generate 100 JPEG image generators that are as
diverse as possible}'', we find that many of the generated images are simply
repeated, and ``100'' is already too large for the LLM to process in one go.
Recent works point out that using only LLMs to generate diverse samples is
inherently challenging~\cite{kandpal2023large,wang2022self}; this is often
referred to as the ``tail phenomena'', where the LLMs tend to generate a large
number of similar samples and only a small number of diverse samples.

\parh{Technical Challenge II: Overhead.}~Real-world fuzzing campaigns require
many input samples to be generated and tested, and the suggested fuzzing
duration is often in the order of days or weeks. This raises a severe concern on
overhead. For example, the cost of using GPT-4 Turbo is estimated to be \$10.00
per 1M tokens, and the cost of using DALL-E is estimated to be \$40.00 per 1,000
images. Given that a single fuzzing campaign may require millions of tokens or
images, the cost of using LLMs can be prohibitively high. 

The time cost of using LLMs is also high, as the generation of a single input
sample may take several seconds or up to twenty seconds, depending on the
complexity of the input format and the quality of the generated sample. This is
not practical for fuzzing, as the fuzzer is expected to generate and test a
large number of input samples in a short time. Suppose each JPEG image takes 10
seconds to generate, and the fuzzer needs to generate 1,000,000 images. This
will take 115 days to complete, which is not practical in real-world fuzzing.

\section{Design}
\label{sec:design}

\begin{figure*}[!htbp]
    \centering
    
    \includegraphics[width=0.9\linewidth]{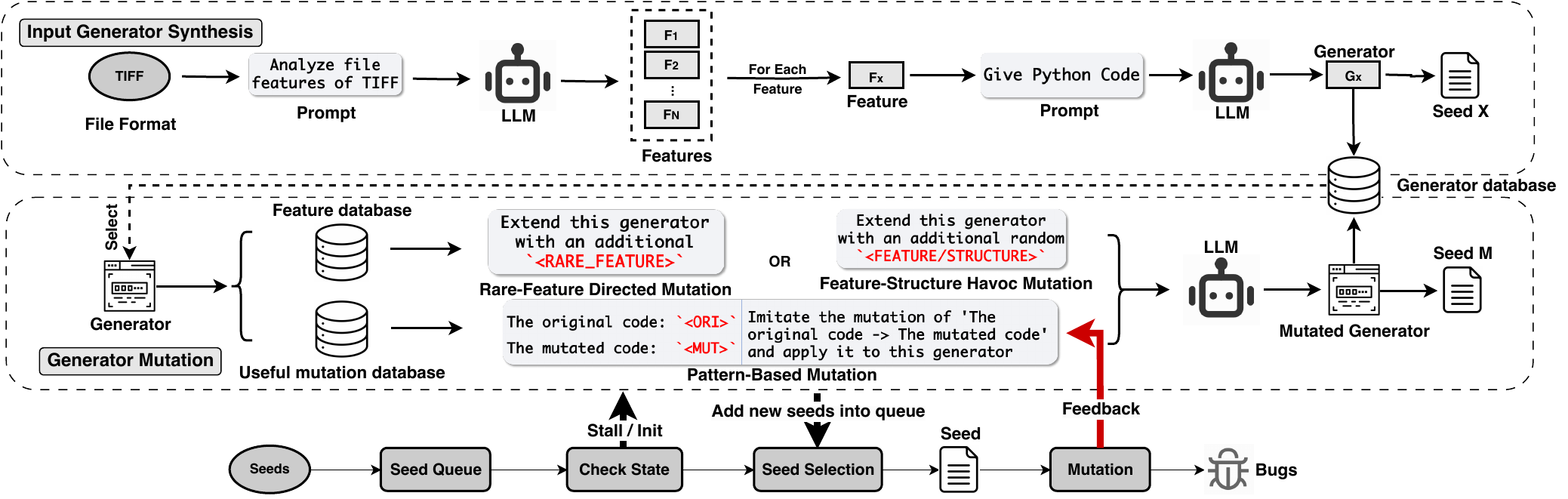}
    \vspace{-7pt}
    \caption{The workflow of \tool.}
    \vspace{-15pt}
    \label{fig:design}
\end{figure*}

In line with challenges noted in \S~\ref{sec:motivation}, we present \tool, a
novel and efficient approach to augment mutation-based fuzzing with LLMs.
\F~\ref{fig:design} illustrates the high-level design of \tool. \tool\ features
a hybrid strategy that combines a ``holistic search'' driven by LLMs and a
``local search'' driven by industrial-quality mutation-based fuzzers. The key
idea is to leverage LLMs to automatically synthesize input generators that are
customized to the specific features, structures, and grammar of the target file
format. We further mutate the synthesized generators to enhance their diversity
(see \S~\ref{sec:synthesis}). By running these generators, \tool\ can obtain
seeds with diverse structures and features. Then, those generated inputs can be
further mutated by traditional mutation-based fuzzers (AFL++) to explore the
input space more effectively (see \S~\ref{sec:mutation}). When \tool\ cannot
identify a new path during the local search, it switches back to the holistic
search to generate new input generators.

\tool\ comprises two core components: \textit{input generator synthesis} and
\textit{input generator mutation}. In input generator synthesis, \tool\ first
analyzes the file features for the target format and synthesizes an input
generator for each feature. At this stage, we obtain some initial, rather simple
generators. In the generator mutation stage, \tool\ aims to produce generators
customized to multiple features or structures simultaneously, which can yield
more complex fuzzing inputs and enhance the generator diversity. We also
evaluate the performance of each generator based on mutation feedback during
fuzzing and extract useful knowledge from the successful generator mutations to
guide future mutation directions. Except for specifying the target file format
(which requires the user to provide), the whole process of \tool\ is automated.

The bottom flowchart in \F~\ref{fig:design} specifies the pipeline of standard
mutation-based fuzzing. We start with initial seeds, add them to the seed queue,
and then select a seed to mutate.  Finally, we mutate
the target seed under the predefined mutation strategy and check if the mutated
inputs trigger bugs.
\tool\ augments the standard pipeline by incorporating the above two LLM-based
components. Before seed selection, we obtain the fuzzing state based on the
fuzzing performance. If it is the first cycle of fuzzing, we perform both input
generator synthesis and input generator mutation to iterate basic features and enrich our initial seed
corpus. If the fuzzing process cannot find a new path for a long time, \tool\
directly mutates input generators, creating more complex generators through feature combinations and
presumably enabling the fuzzing campaign to escape from ``local optima.'' 

%


\parh{Synergy Effect.}~We highlight that \tool\ features a synergistic effect
when combined with mutation-based fuzzers. LLMs are knowledgeable about the
grammatical information of various input formats, yet they are less capable of
generating those non-textual inputs directly. On the other hand, mutation-based
fuzzers are good at performing fine-grained, byte-level mutations and exploring
the input space systematically at low cost. However, conventional mutation-based
fuzzers often lack the grammatical knowledge to generate high-quality input
samples, and they often lack the ``big picture'' to progressively explore the
input space. A good synergy between LLMs and mutation-based fuzzers can be
achieved, where LLMs excel at synthesizing input generators and enabling
escape from local optima, and mutation-based fuzzers excel at deeply exploring
the local input space. This alleviates the limitations of both LLMs and
conventional mutation-based fuzzers, and achieves better performance than using
either of them alone; see evaluations in \S~\ref{sec:evaluation}.

\parh{Addressing Technical Challenge I.}~Challenge I concerns the lack of
diversity in LLM-generated outputs. As aforementioned, LLMs are inherently prone
to the ``tail phenomena'' and often generate outputs that are repeated or very
similar to each other. Rather than directly asking LLMs to produce ``diverse''
generators, \tool\ first analyzes the possible features of a target file format,
and then uses LLMs to synthesize input generators tailored to specific
features/structures of the target file format. We also propose a set of
strategies to extend and mutate the synthesized generators.

\parh{Addressing Technical Challenge II.}~Challenge II concerns the high cost of
LLM usage. We address this challenge in a principled manner, where 
we only invoke LLMs to generate new input generators when needed.
Holistically, LLMs are only invoked when the local search (conducted by AFL++)
cannot identify new edges. This largely reduces the cost of LLM usage, from
15.16\$ (our LLM-baseline setting; see comparisons in \S~\ref{sec:evaluation})
to 0.124\$, thus making \tool\ practical and cost-friendly in real-world fuzzing
campaigns.

\parh{Application Scope.}~\tool\ is designed to be general-purpose and
applicable to a wide range of input formats. We have evaluated \tool\ on a
variety of input formats, including JPEG images, TIFF images, MP4 videos, and 31
other formats. The design of \tool\ is not specific to any
particular input format, and it can be easily extended to support new input
formats. With modern LLMs increasingly capable of gaining complex grammatical
knowledge and coping with advanced data types (e.g., videos and audio), we are
positive that \tool\ can be used to augment mutation-based fuzzing for a wide
range of input formats. We leave the exploration of those advanced data types to
future work.







\subsection{Input Generator Synthesis}
\label{sec:synthesis}


\parh{When to Use.}~Before entering the formal fuzzing loop, \tool\ first
obtains the target input file format from the user (e.g., ``TIFF''; this
is the \textit{only} information required), extracts its features, and
synthesizes the corresponding generators. Then, it runs these generators previously-synthesized by the LLM to
produce new diverse seeds and adds them to the seed queue. Note
that if a software accepts multiple input file formats, \tool analyzes each
format individually to obtain the corresponding generators.

\parh{Design Consideration: Features vs. Structures.}~To describe a file, two
main aspects can be considered: \textit{feature} and \textit{structure}. ``File
feature'' refers to attributes or characteristics that can provide external
details about the file. ``File structure'' refers to the way data within a file
is organized and formatted, providing internal details about how data is
arranged within the file. Using file structure to describe a desired file input
is a more straightforward approach. However, there is a gap between specifying
structure and preparing the generator Python code. The document of Python file
libraries often lacks details on how to write code to yield a specific file
structure. In Python, constructing a file with a specific structure is not like
building with blocks, where one chunk can be added at a time; instead, the file
is often constructed from a more holistic perspective. This makes it difficult
for an LLM to understand and use the libraries to achieve certain
structures. Also, relations among chunks can be complex and have
intricate dependencies. 
Our tentative study shows that creating generators based on structures has a
high failure rate, consuming much time and negatively affecting fuzzing throughput.

We find that relevant Python file libraries often provide APIs to implement
features for specific file formats, such as the compression flag in
\texttt{libtiff} for storing TIFF files. In these cases, features and code have
a direct map, as the document of these libraries includes corresponding
descriptions. LLMs can learn from this information, making file features easy
for them to understand. Consequently, the transformation from features to a
Python generator code is straightforward. Indeed, since file features also
encompass structural descriptions and complex constraints, generating input with
a specific feature must adhere to the grammar and structural constraints.
Therefore, we use file features to synthesize generators.

\begin{algorithm}[h]
\scriptsize
\caption{Generator Procedure}\label{alg:overviewSynthesis}
\KwIn{The target file format, $target\_format$.}
\KwOut{A set of (generator, feature description), $G$.}
$prompt \gets construct\_prompt(target\_format)$ \tcp*[f]{Based on \F~\ref{fig:featurePrompt}}\\
$features \gets LLM(prompt)$\\
\For{$f$ in $features$}{
    $g \gets generator\_generation(target\_format, target\_feature)$\\ \tcp*[f]{Based on Alg. \ref{alg:programGeneration}}\\
    \If{$g \neq none$}{
        $seeds \gets run(g)$\\
        $add\_to\_queue(seeds)$\\
        $G \gets (g, f)$\\
    }
}
\end{algorithm}

\parh{Overview.}~Input generator synthesis has two steps: given a file
format, we first perform feature analysis to identify all possible features.
Then, for each feature, we ask LLMs to synthesize a generator that produces
an input with the target feature. As shown in \A~\ref{alg:overviewSynthesis},
given a file format, \tool\ constructs a prompt and asks the LLM for the
corresponding features (lines 1-2). For each feature, \tool\ leverages an LLM to
create a generator for it (lines 3-4). Then, \tool\ runs the generators to
obtain seeds with various features and adds these seeds to the seed queue in
fuzzing (lines 6-8).

\begin{figure}[!htbp]
    \vspace{-5pt}
    \centering
    \includegraphics[width=0.95\linewidth]{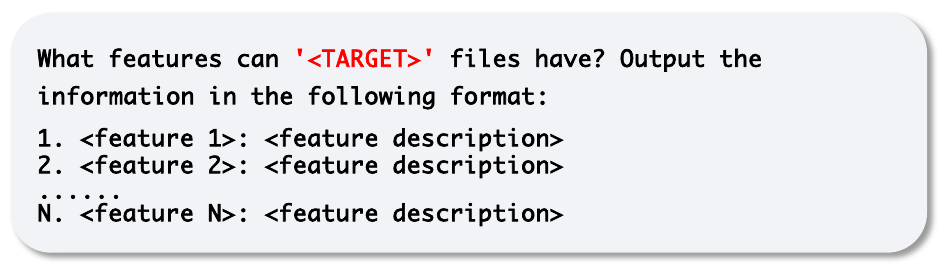}
    \vspace{-5pt}
    \caption{The prompt used to analyze the features of a specific program.}
    \vspace{-10pt}
    \label{fig:featurePrompt}
\end{figure}

\parh{Feature Analysis.}~As there are many feature descriptions in the document
of Python file libraries, the LLM can synthesize a generator producing a file
with the specific feature. To obtain the features for a given file format, we
instruct the LLM to summarize the possible features. The prompt is shown in
\F~\ref{fig:featurePrompt}. For example, when applying this prompt to extract
the features of TIFF files, the output might include: \textit{1. Lossless
compression: TIFF files support ... 2. Multiple layers: ...}. We do not limit
the number of features, as different file formats have varying ranges of
features. We aim to capture common features at this step, and explore unusual
features in \S~\ref{sec:mutation}.

\parh{Generator Synthesis.}
After we obtain the features for a specific file format, we synthesize a
generator for each feature.
For the generator, we have two requirements: (i) it should be written in Python,
and (ii) it should be executable. Obtaining a Python generator is
straightforward for LLMs. However, there are several challenges for generators
to run smoothly. First, as seed construction may rely on certain Python
libraries, it is common to encounter the \textit{ModuleNotFound} problem
(Challenge I). Thus, we use the LLM to analyze the error information to
automatically identify the required libraries and install them. Second, as LLMs
cannot ensure the validity of generated codes, the code generated by an LLM may
contain some bugs (Challenge II). We propose an algorithm to automatically debug
the generated code.

\begin{algorithm}[!t]
\scriptsize
\caption{Generator Synthesis Algorithm}\label{alg:programGeneration}
\KwIn{The target file format, $target\_format$. The target file feature, $target\_feature$.}
\KwOut{A valid generator that can generate a file with specific features, $g$.}
$init\_cnt \gets 0$\\
\While{$init\_cnt < INIT\_MAX$}{
    $dialogue \gets [\ ]$\\
    $prompt \gets construct\_prompt(target\_format, target\_feature)$ \\ \tcp*[f]{Based on \F~\ref{fig:createGenPrompt}}\\
    $dialogue.append(prompt)$\\
    $g \gets LLM(dialogue)$\\
    $status, msg \gets exec(g)$\\
    $debug\_cnt \gets 0$\\
    \While{$debug\_cnt < DEBUG\_MAX$}{
        \If{$status == SUCCESS$}{
            $ruturn\ g$
        }
        $error\_info \gets get\_msg(msg)$\\

        \While{$TRUE$}{
            \If{$"ModuleNotFoundError"$ not in $error\_info$}{
                $break$
            }

            $library\_prompt \gets construct\_prompt(error\_info)$ \\ \tcp*[f]{Based on \F~\ref{fig:libraryPrompt}}\\
            $relied\_library \gets LLM(library\_prompt)$\\
            $flag, g = automatic\_installation(relied\_library)$\\

            \If{$flag == 0$}{ \tcp*[f]{Failed to install the library}\\
                $ruturn\ None$
            }

            $status, msg \gets exec(g)$\\
            \If{$status == SUCCESS$}{
                $ruturn\ g$
            }\Else{
                $error\_info \gets get\_msg(msg)$\\
            }
        }

        $dialogue.append(g)$\\
        $dialogue.append(error\_info + ``Regenerate")$\\
        $G \gets LLM(dialogue)$\\
        $status, msg \gets exec(g)$\\
        $debug\_cnt \gets debug\_cnt + 1$\\
    }
    $init\_cnt \gets init\_cnt + 1$\\
}
\end{algorithm}

\begin{figure}[!htbp]
    \centering
    \vspace{-5pt}
    \includegraphics[width=0.95\linewidth]{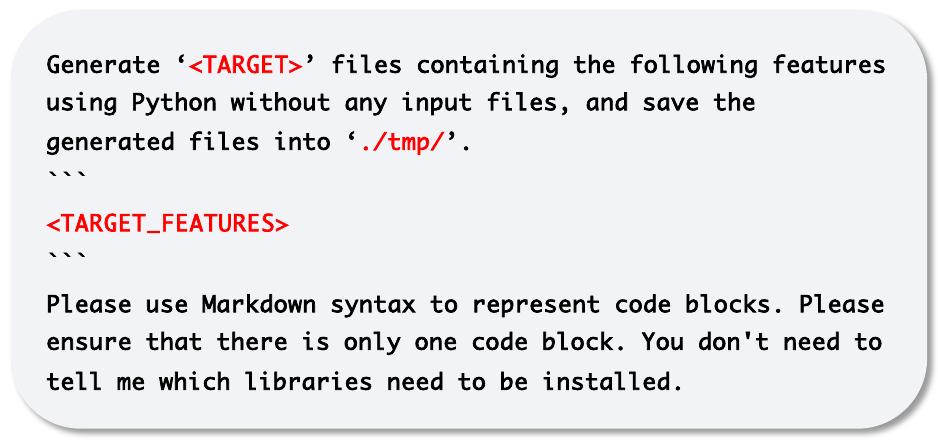}
    \vspace{-5pt}
    \caption{The prompt for developing a generator from a specific
    feature.}
    \vspace{-15pt}
    \label{fig:createGenPrompt}
\end{figure}

As in \A~\ref{alg:programGeneration}, the synthesis algorithm first constructs
an initial generator based on the target file format and the desired feature,
and runs it to obtain the execution status (lines 3-7). The prompt template used
is shown in \F~\ref{fig:createGenPrompt}. If the execution fails, the specific
error message is extracted (line 12).

\begin{figure}[!htbp]
    \centering
    \includegraphics[width=0.95\linewidth]{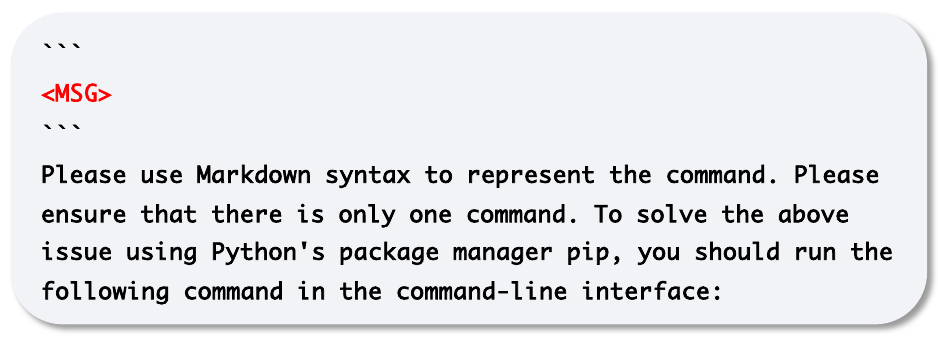}
    \vspace{-5pt}
    \caption{The prompt for extracting the required library.}
    \vspace{-10pt}
    \label{fig:libraryPrompt}
\end{figure}

If the error involves a missing module, the algorithm attempts to install the
required library. This process involves constructing a prompt based on the error
information, using the LLM to identify the missing library, and then attempting
an automated installation (lines 14-18). The prompt template is in
\F~\ref{fig:libraryPrompt}. If the installation is successful, the generator is
re-executed, and if the execution status is \textit{SUCCESS}, the valid
generator is returned (lines 22-23). Otherwise, the error handling loop
continues.

After resolving the library dependency issue, the error information is fed back into the LLM to regenerate a program that can resolve the current error (lines 26-28).
If debugging up to $DEBUG\_MAX$ times still fails to produce a valid generator, the algorithm attempts to generate a new initial generator (back to line 3).
This is crucial because, due to the stochastic nature of LLMs, the same prompt can yield generators of varying quality, helping to avoid getting stuck in ``local minima'' (lines 9-12 and lines 26-30).
Based on our preliminary exploration, we set $INIT\_MAX$ to 2 and $DEBUG\_MAX$ to 3 to balance the trade-off between the quality of the generated generator and the time cost.



\subsection{Generator Mutation}
\label{sec:mutation}

The generators obtained from \S~\ref{sec:synthesis} generate a seed with a
single specific feature, which often covers a small part of the feature space.
To effectively utilize the mutation feedback information from fuzzing and cover a larger feature space, we take into account more complex features using the following three mutation strategies:
    \textbf{Rare-Feature Directed Mutation:} We incorporate historical
    information—specifically, the features that have already been covered by the generators—into the prompt to guide the LLM in extracting unanalyzed
    features, focusing specifically on adding these rare features to the
    existing generators.
    \textbf{Feature-Structure Havoc Mutation:} We add a random
    feature/structure to the existing generators, aiming to unleash the
    potential upper bound capability of the LLM.
    \textbf{Pattern-Based Mutation:} As different features may exert
    varying influences on the target program, we leverage historical information
    to extract useful features and retain them by combining them with other
    features. Thus, we use the feedback information from the fuzzing process to guide the generator mutations.

\parh{When to Use.}
In the fuzzing process, generator mutation is executed in two specific
situations. First, when \tool\ initially enters the fuzzing loop, we use
rare-feature directed mutation to enrich the variety of features in the
initial seeds. Second, when fuzzing fails to find new paths within a set time
limit (likely trapped in a local optimum), we use feature-structure havoc
mutation and pattern-based mutation to construct seeds with different features
or structures, which can help fuzzing explore other code regions.

\parh{Overview.}~When fuzzing stalls or initializes, \tool uses
generator mutation to generate more complex inputs. The algorithm is in
\A~\ref{alg:overviewMutation}. \tool obtains the current fuzzing state (line 1).
If it is initialization, \tool performs rare-feature directed mutation. To do
so, \tool asks LLMs to extract the unanalyzed features based on historical
information (lines 3-4). For each unanalyzed feature, \tool randomly selects a
generator and asks LLMs to incorporate the unanalyzed feature into it to create
new inputs (lines 5-11).

If it is a stall, \tool performs feature-structure havoc mutation or
pattern-based mutation. \tool randomly selects a generator from a database
containing all executable generators (line 13). It then randomly chooses a
mutator to apply to this selected generator (line 14). Next, \tool constructs a
prompt based on the chosen generator and mutator (lines 15-19). Finally, \tool
retrieves a mutated generator from the LLM, runs it to obtain a new seed, and
adds this seed to the queue for further mutation (lines 20-23).

\begin{algorithm}[!t]
\scriptsize
\caption{Generator Mutation Algorithm}\label{alg:overviewMutation}
\KwIn{The target file format, $format$.}
\KwOut{A generator, $g_{m}$.}

$state \gets get\_fuzz\_state()$\\
\If{$state == init$}{
    $prompt \gets construct\_prompt(target\_format)$ \tcp*[f]{Based on \F~\ref{fig:rareFeature}}\\
    $features \gets LLM(prompt)$\\
    \For{$f$ in $features$}{
        $g \gets generator\_select()$\\
        $prompt \gets construct\_prompt(format, g, f)$ \tcp*[f]{Based on \F~\ref{fig:rareMutation}}\\
        $g_{m} \gets LLM(prompt)$\\
        $g_{m} \gets self\_debug(g_{m})$ \tcp*[f]{Reuse the code lines 9 - 30 in \A~\ref{alg:programGeneration}}\\
        $seeds \gets run(g_{m})$\\
        $add\_to\_queue(seeds)$
    }
}

\If{$state == stall$}{
    $g \gets generator\_select()$\\
    $mutator \gets mutator\_choose()$\\
    \If{$mutator == feature$ or $mutator == structure$}{
        $prompt \gets construct\_prompt(format, g, mutator)$ \tcp*[f]{Based on \F~\ref{fig:randomMutation}}\\
    }
    \ElseIf{$mutator == pattern$}{
        $example \gets pre\_mutation\_select()$\\
        $prompt \gets construct\_prompt(g, example)$ \tcp*[f]{Based on \F~\ref{fig:patternMutation}}\\
    }

    $g_{m} \gets LLM(prompt)$\\
    $g_{m} \gets self\_debug(g_{m})$ \tcp*[f]{Reuse the code lines 9 - 30 in \A~\ref{alg:programGeneration}}\\
    $seeds \gets run(g_{m})$\\
    $add\_to\_queue(seeds)$
}
\end{algorithm}

\parh{Rare-Feature Directed Mutation.}
To improve the comprehensiveness of our testing, it is essential to cover rare
features that the generators from \S~\ref{sec:synthesis} may have overlooked.
Our tentative study shows LLMs cannot often identify all relevant features of a
file format in a single request. Typically, they provide around ten features at
a time but often neglect rare features and cannot generate them directly.

\begin{figure}[!htbp]
    \centering
    \vspace{-5pt}
    \includegraphics[width=0.92\linewidth]{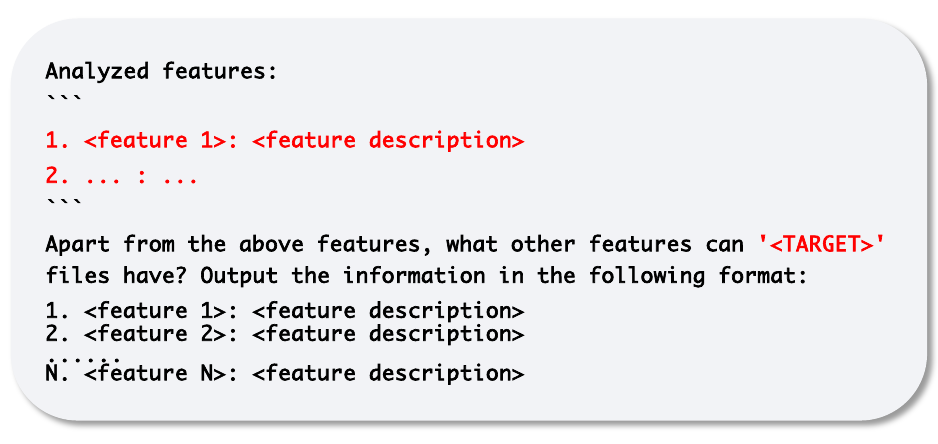}
    \vspace{-5pt}
    \caption{The prompt for rare feature extraction.}
    \vspace{-15pt}
    \label{fig:rareFeature}
\end{figure}

To achieve rare feature mutation, we maintain a feature database that collects
analyzed features as described in \S~\ref{sec:synthesis}. Once a feature has
been analyzed to synthesize a generator, its name, and corresponding description
are added to the feature database. We then incorporate these analyzed features
into a prompt and ask the LLM to identify other unexplored features, as
illustrated in \F~\ref{fig:rareFeature}.

\begin{figure}[!htbp]
    \centering
    \vspace{-5pt}
    \includegraphics[width=0.9\linewidth]{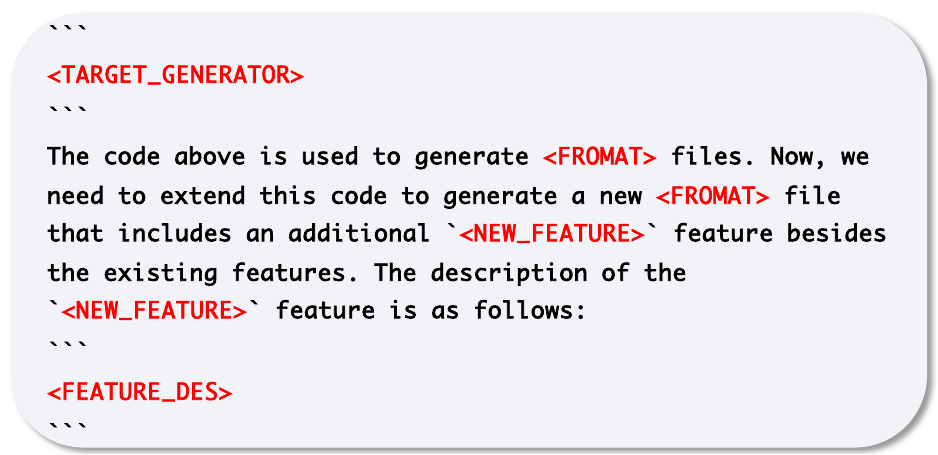}
    \vspace{-5pt}
    \caption{The prompt for rare feature mutation.}
    \vspace{-10pt}
    \label{fig:rareMutation}
\end{figure}

At the same time, we store all the synthesized generators in the generator
database, and we randomly select a generator from this database. Afterward, we
ask the LLM to mutate the selected generator to produce a file that includes an
additional rare feature alongside the existing ones. The prompt for this step is
shown in \F~\ref{fig:rareMutation}. Finally, we run the mutated generator,
obtain new seeds with multiple (newly-added) features, and add this seed to the
seed queue. Given that this method requires putting all previously analyzed
feature descriptions into a prompt, we only use this strategy the first time
fuzzing enters the loop to reduce token overhead.

\parh{Feature-Structure Havoc Mutation.}
Although the LLM is powerful, its output can sometimes be unstable. To explore
the full potential of the LLM, we ask it to randomly mutate the current
generator to produce a file that includes an additional feature or structure
alongside the existing features. The prompt is shown in
\F~\ref{fig:randomMutation}.
Since a generator can be mutated multiple times, it is possible to generate a
file with many features or structures. The randomness of the LLM may introduce
rare features that cannot be discovered through directed rare-feature mutation.
While rare-feature directed mutation typically generates a file with two
features, feature-structure havoc mutation can produce a file with more than two
features. This allows for the construction of more complex generators, enabling
us to explore a deeper feature space.

\parh{Pattern-Based Mutation.}
Given that different features may exert varying influences on the target
program, we propose pattern-based mutation. This approach uses historical
information to extract useful features, which are then accentuated by
integrating them with other features. Feedback from the fuzzing process
effectively highlights which mutations result in more useful generators (i.e.,
those capable of discovering new edges). By analyzing this feedback, we can ask
LLMs to learn these mutation strategies, thus guiding and optimizing future
mutation directions.

The feature space of a file format is often vast, and not every unique feature
triggers distinct processing logic in the target program. Iterating through all possibilities is inefficient; instead, we focus on the
features that the target program is interested in, namely ``program-relevant''
features. Seeds with program-relevant features will be processed differently by
the target program. If a seed obtained from a mutated generator discovers a new
path, we infer that this seed contains program-relevant features. Consequently,
we consider the $<$original generator, mutated generator$>$ tuple to contain
useful information and incorporate it into our useful pattern database.

\begin{figure}[!htbp]
    \vspace{-5pt}
    \centering
    \includegraphics[width=0.9\linewidth]{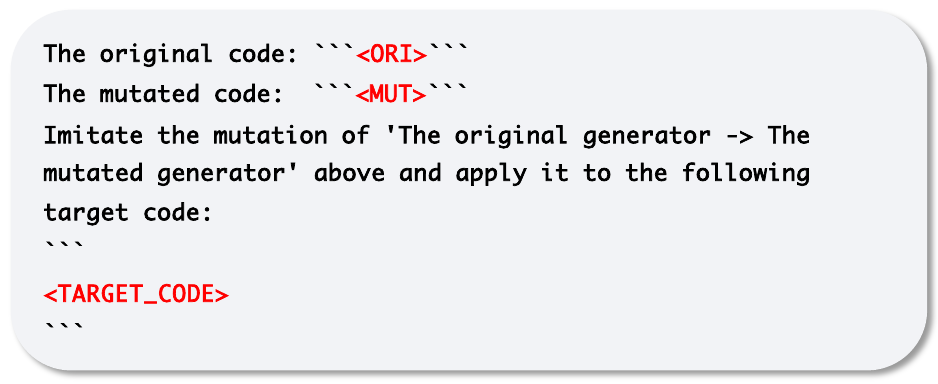}
    \vspace{-5pt}
    \caption{The prompt for pattern-based mutation.}
    \vspace{-15pt}
    \label{fig:patternMutation}
\end{figure}

To reuse effective mutation strategies, we employ LLMs to learn the mutation
patterns from the mutation generator tuples and apply these patterns to other
generators. The prompt is shown in \F~\ref{fig:patternMutation}. By doing so, we
aim to generate seeds with a richer variety of program-relevant features.
In our implementation, we trace the performance of each generated
seed during fuzzing. If a useful seed (i.e., one that discovers new paths) is
produced through generator mutation, we consider this mutation useful for this
program. Therefore, we add both the mutated and original generators to the prompt in
\F~\ref{fig:patternMutation} and apply this mutation strategy to
other generator mutations.

\section{Evaluation}
\label{sec:evaluation}
\tool is built upon AFL++, enabling integration of our method with other
existing techniques. To ensure the accuracy and fairness of our results, we
conducted experiments on three testing platforms: UNIFUZZ, MAGMA, and FuzzBench.
The experiments were carried out on three systems running Ubuntu 22.04, each
equipped with 64 cores (Intel(R) Xeon(R) Gold 6444Y CPU) and 256GB memory. We
study the following research questions (RQs): \textbf{RQ1:} Can the tool
outperform SOTA in terms of code coverage and the number of unique bugs?
\textbf{RQ2:} Can \tool's performance surpass that of structure-aware fuzzers?
\textbf{RQ3:} How many tokens will be consumed when fuzzing a program for 24
hours? \textbf{RQ4:} Which part of \tool contributes the most?

\subsection{Code Coverage and Unique Bugs}

Code coverage and unique bugs are common metrics for evaluating fuzzers. To ensure fairness and reproducibility, we conduct our experiments on UNIFUZZ, MAGMA, and FuzzBench. All runtime settings, including the initial seeds, follow the default configuration.

\subsubsection{Experiments on UNIFUZZ}
\label{sec:unifuzz}
UNIFUZZ is an open-source and metrics-driven platform designed for the holistic and fair evaluation of fuzzers. 

\noindent\textbf{Compared Fuzzers.}
AFL++'s performance demonstrates that the implementation
significantly affects testing efficiency. 
To avoid the influence of implementation, we have developed \tool based on AFL++, integrating it as a 
mode within AFL++. Since AFL++ already incorporates many SOTA
fuzzers, this allows for easy and fair comparisons between \tool and other
fuzzers implemented in AFL++. As the range of features incorporated into AFL++
would exceed the scope of this paper, we compared \tool with the four most widely
used configurations of AFL++. \textbf{AFL++(cmplog)}: enables REDQUEEN mutator.
\textbf{AFL++(mopt)}: enables MOPT mutator. \textbf{AFL++(fast)}: enables
AFLFast seed scheduling. \textbf{AFL++(rare)}: prioritizes seeds that are rarely
covered by other seeds. 
For \tool, we enable cmplog mode to facilitate efficient low-level mutation.


\noindent\textbf{Programs Selection.}
Since \tool is designed for testing programs with non-textual inputs, we selected programs that meet this criterion. The target programs, listed in \T~\ref{tab:allPrograms}, include 10 programs with over 20 different input format types.

\begin{table*}[t]
    \small
    \footnotesize
    \centering
    \vspace{-5pt}
    \caption{Average code coverage and the total unique crashes found by \tool
    with GPT-3.5/GPT-4 and 6 compared fuzzers.}
    \vspace{-8pt}
    \label{tab:motivation}
    \resizebox{0.81\linewidth}{!}{
    \begin{threeparttable}
    \begin{tabular}{c | c | c | c | c | c | c | c | c | c | c | c | c }
    \toprule
    \multirow{2}{*}{Programs} & \multicolumn{2}{c|}{\textbf{\textsc{G$^2$Fuzz}(GPT-3.5)\tnote{1}}}  & \multicolumn{2}{c|}{\textbf{\textsc{G$^2$Fuzz}(GPT-4)\tnote{2}}}  & \multicolumn{2}{c|}{\textbf{AFL++(cmplog)}}& \multicolumn{2}{c|}{\textbf{AFL++(fast)}}  & \multicolumn{2}{c|}{\textbf{AFL++(mopt)}} & \multicolumn{2}{c}{\textbf{AFL++(rare)}}\\
    & Cov. & \#Bug & Cov. & \#Bug & Cov. & \#Bug & Cov. & \#Bug & Cov. & \#Bug & Cov. & \#Bug \\
    \midrule

exiv2 & 5,099 & \textbf{31} & \textbf{5,171} & 28 & 4,965 & 26 & 3,776 & 5 & 3,758 & 12 & 3,851 & 11 \\
ffmpeg & 31,706 & 0 & \textbf{34,218} & \textbf{1} & 22,099 & 0 & 17,380 & 0 & 15,566 & 0 & 14,613 & 0 \\
flvmeta & 228 & 4 & 228 & 4 & 228 & 4 & 228 & 4 & 228 & 4 & 228 & 4 \\
gdk & \textbf{2,958} & \textbf{6} & 2,327 & 2 & 2,172 & 5 & 2,093 & 4 & 2,082 & 2 & 1,991 & 4 \\
imginfo & 2,839 & 0 & \textbf{3,825} & 0 & 2,189 & 0 & 1,998 & 0 & 2,004 & 0 & 1,976 & 0 \\
jhead & 315 & 13 & \textbf{491} & \textbf{23} & 445 & 21 & 195 & 3 & 195 & 3 & 195 & 4 \\
mp3gain & 921 & 11 & 921 & \textbf{12} & \textbf{923} & 11 & 900 & 10 & 899 & 8 & 891 & 10 \\
mp42aac & 2,067 & \textbf{17} & \textbf{2,700} & 14 & 2,091 & 13 & 1,178 & 6 & 1,157 & 3 & 1,135 & 2 \\
pdftotext & \textbf{8,265} & 43 & 7,921 & \textbf{49} & 7,434 & 25 & 6,483 & 25 & 6,376 & 28 & 11,257 & 28 \\
tiffsplit & 1,817 & 7 & \textbf{1,840} & \textbf{10} & 1,659 & 6 & 1,619 & 9 & 1,644 & 9 & 1,626 & 7 \\
    
    \bottomrule
    \end{tabular}
    
    \begin{tablenotes}
    \item[1] \textbf{\textsc{G$^2$Fuzz}(GPT-3.5)}: \tool based on GPT-3.5.
    \item[2] \textbf{\textsc{G$^2$Fuzz}(GPT-4)}: \tool based on GPT-4.
    \end{tablenotes}
    \end{threeparttable}
    }
    \vspace{-10pt}
    \label{tab:unifuzzEdge}
\end{table*} 

\noindent\textbf{Experiment Results.}~Table~\ref{tab:unifuzzEdge} shows the edge coverage
achieved by eight fuzzers. 
\textsc{G$^2$Fuzz}(GPT-4) discovers a total of
59,642 edges, which is 15,437 more than the best baseline fuzzer, AFL++(cmplog).
\textsc{G$^2$Fuzz}(GPT-4) and \textsc{G$^2$Fuzz}(GPT-3.5) achieve the highest
performance on 9 out of 10 programs, while AFL++(cmplog) excels on one program.
Furthermore, \textsc{G$^2$Fuzz}(GPT-4) is able to discover more unique bugs than the other
baseline fuzzers. Specifically, \textsc{G$^2$Fuzz}(GPT-4) finds 143 unique bugs,
which is 32 more than the best-performing comparison fuzzer, AFL++(cmplog).

\begin{figure}[!htbp] 
    \centering
    \includegraphics[width=0.475\textwidth]{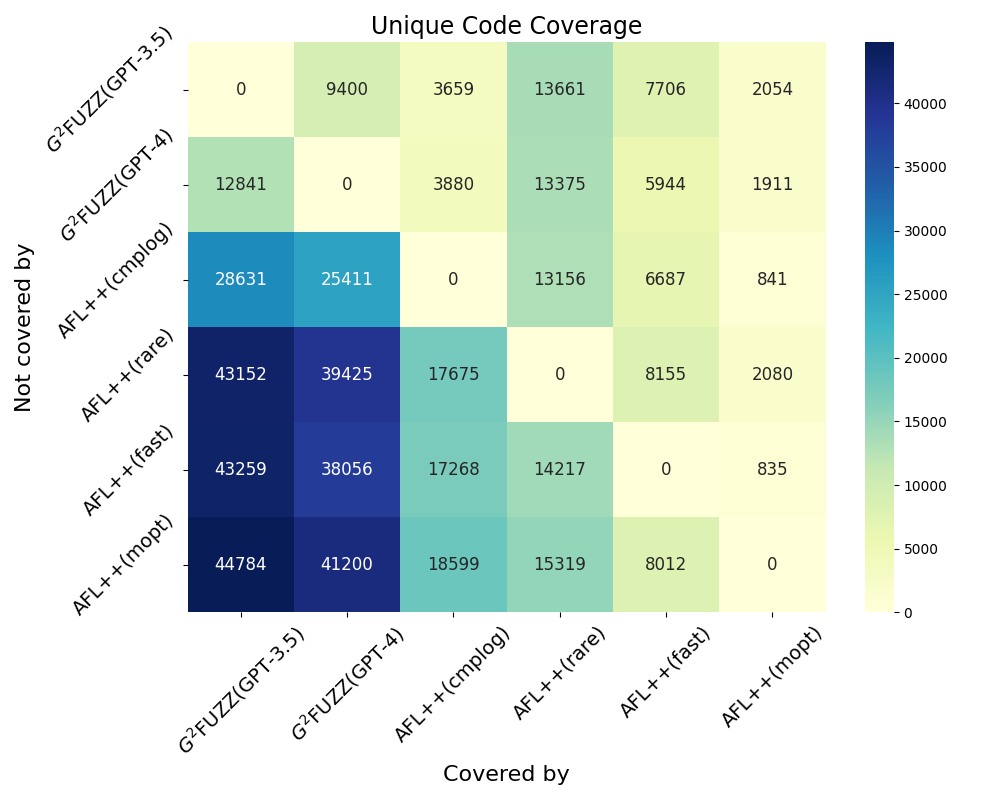}
    \vspace{-20pt}
    \caption{Pairwise unique code coverage across all programs. Each cell represents the number of code branches covered by the fuzzer of the column but not by the fuzzer of the row.}
    \vspace{-15pt}
    \label{fig:uniqueEdgeCov}
\end{figure}

Moreover, we also calculated the pairwise unique code coverage. We summed up the
unique code coverage for each pair of fuzzers across all programs, resulting in
\F~\ref{fig:uniqueEdgeCov}. \textsc{G$^2$Fuzz}(GPT-4) and
\textsc{G$^2$Fuzz}(GPT-3.5) are quite similar, and both are able to identify
more unique code coverage than the remaining four fuzzers. This result
demonstrates that with the assistance of LLMs, the inputs we generated, which
possess complex features, are capable of triggering more intricate program
logic. Consequently, this leads to the discovery of a higher amount of unique
code coverage.

\tool incorporates additional steps into AFL++, such as generator synthesis and
execution. To assess the impact of these steps on fuzzing throughput (i.e.,
execution speed), we measure the total number of executions for each fuzzer.
\F~\ref{fig:throughput} presents the total number of executions performed by all
programs that the fuzzers run after 24 hours. The results indicate that \tool's
throughput does not significantly decrease compared to other fuzzers.
We also fuzzed certain programs for 48 hours to observe \tool's performance in the later stages of fuzzing. During the 24 to 48-hour period, \tool discovered 32, 5, 1, 33, and 89 new edges in imginfo, jhead, mp3gain, mp42aac, and tiffsplit, respectively.

Furthermore, we evaluated the token costs of LLMs for fuzzing. \tool minimizes token usage by reducing reliance on LLMs for mutation. GPT-3.5 incurs costs of less than 0.2\$, and GPT-4 less than 13\$ for 24 hours of fuzzing. More details can be found in Appendix~\ref{sec:TokenCost}. Additionally, the ablation study shows that both components of \tool are effective, with generator synthesis and mutation contributing 82,001 and 141,340 new paths, respectively. LLM-only approaches struggle, highlighting the necessity of integration. For more details, please refer to Appendix~\ref{sec:Ablation}.

\subsubsection{Experiments on FuzzBench}
To conduct more comprehensive experiments, we also evaluate \tool using FuzzBench.

\noindent\textbf{Experiment Setup.}
FuzzBench builds a Docker image for each fuzzer-benchmark pair to run the experiments. However, the resulting container does not have an internet connection. Therefore, we modify the condition to run the LLM-generation algorithm from `stall/init' to `init' to create a variant of \tool. Specifically, we run the input generator synthesis and generator mutation when the fuzzing process first enters the fuzzing loop.
This allows us to run the algorithm in advance and upload the results into the container built by FuzzBench. Consequently, we can conduct the experiments without needing an internet connection. 
To mitigate randomness in LLM generation, we conduct three epochs to obtain three sets of seeds from the LLM-generation algorithm. Therefore, we need to perform three sets of experiments.

\noindent\textbf{Programs Selection.}
Since \tool is suitable only for non-textual inputs, we excluded programs with textual inputs, leaving 11 programs for testing \tool, as listed in \T~\ref{tab:allPrograms}.


\noindent\textbf{Metric.} 
We choose the average rank of fuzzers as our evaluation metric to assess each fuzzer's performance across multiple benchmarks. By ranking the fuzzers on each benchmark according to their median reached code coverage, with lower values indicating better performance, we can derive an overall understanding of their effectiveness. 

\begin{figure*}[t] 
    \centering
    \begin{minipage}[t]{0.24\textwidth}
        \centering
        \includegraphics[width=\textwidth]{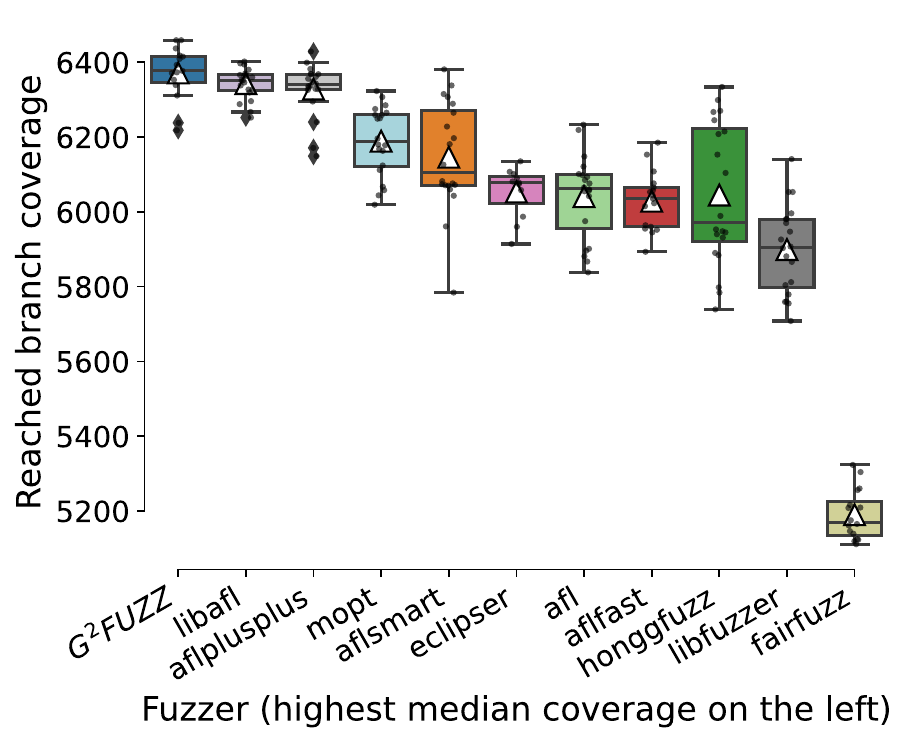}
        \vspace{-17pt}
        \subcaption{bloaty}
        \vspace{-10pt}
    \end{minipage}
    \begin{minipage}[t]{0.24\textwidth}
        \centering
        \includegraphics[width=\textwidth]{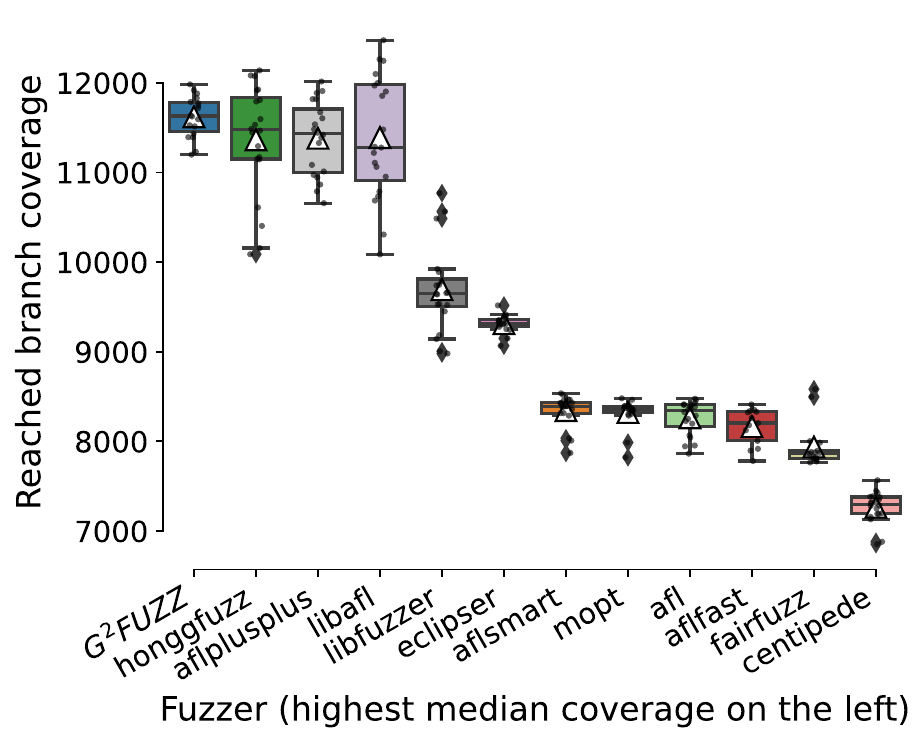}
        \vspace{-17pt}
        \subcaption{freetype2}
        \vspace{-10pt}
    \end{minipage}
    \begin{minipage}[t]{0.24\textwidth}
        \centering
        \includegraphics[width=\textwidth]{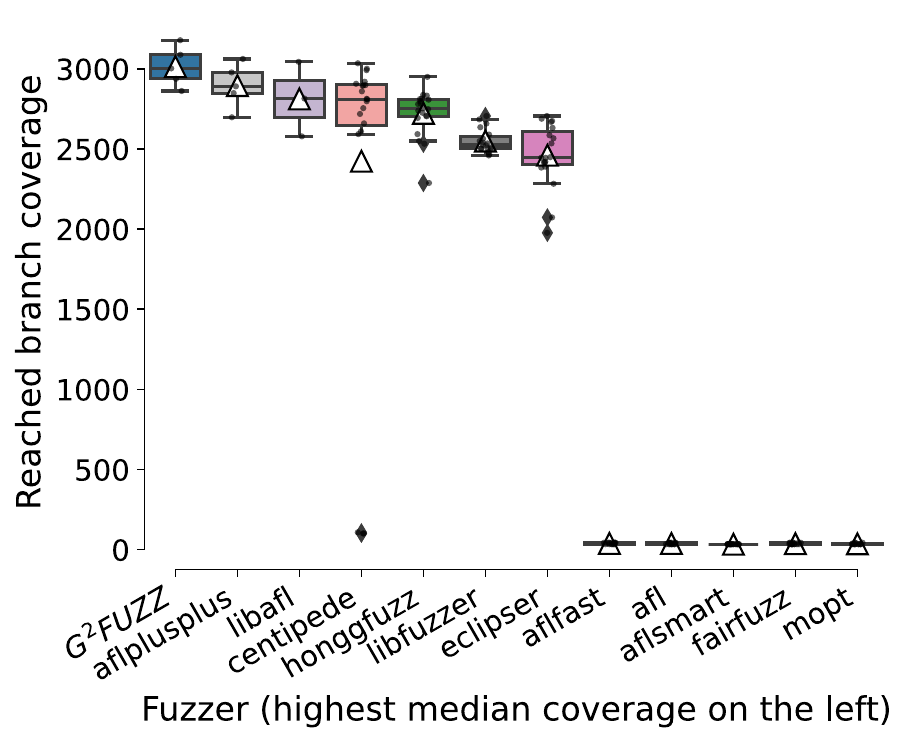}
        \vspace{-17pt}
        \subcaption{libpcap}
        \vspace{-10pt}
    \end{minipage}
    \begin{minipage}[t]{0.24\textwidth}
        \centering
        \includegraphics[width=\textwidth]{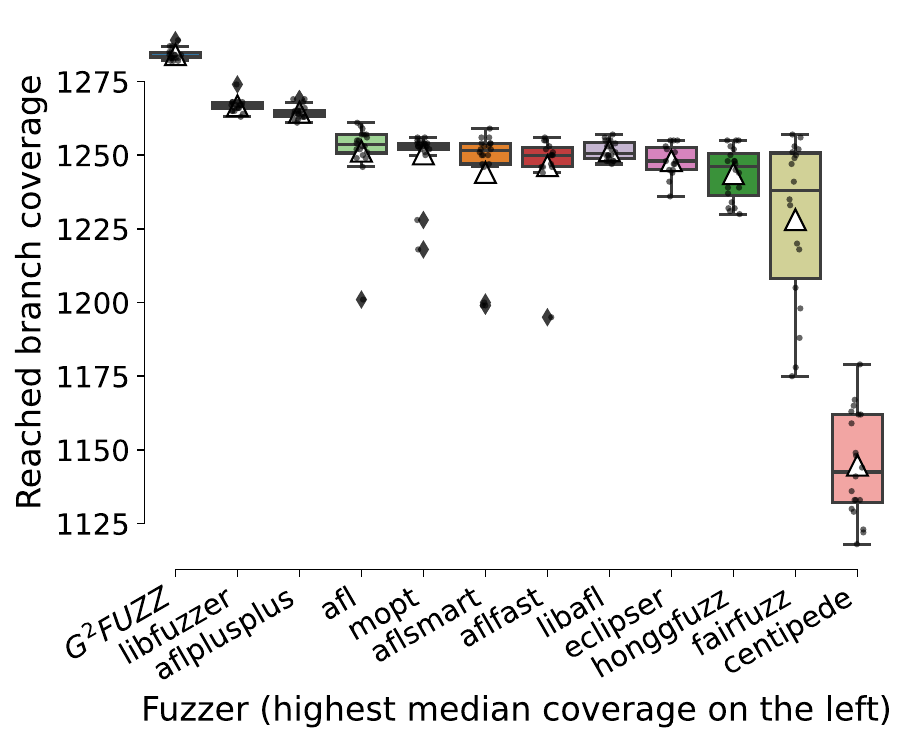}
        \vspace{-17pt}
        \subcaption{vorbis}
        \vspace{-10pt}
    \end{minipage}

    \caption{Code coverage distributions achieved in a FuzzBench experiment. Due
    to space constraints, we only present the results on four programs. For
    other experimental results, refer to \F~\ref{fig:fuzzbenchOthers}.}
    \vspace{-13pt}
    \label{fig:fuzzbenchAll}
\end{figure*}

\noindent\textbf{Experiment Results.}
As shown in \T~\ref{tab:fuzzbenchRank}, the performance of \tool remains stable across different experimental groups.
\tool achieves the best ranks in all three groups, which are 2.09, 2.18, and 2.18. 
The second-best fuzzer, AFL++, achieves ranks of 2.73, 2.73, and 2.91 in the respective groups. 
In Figure~\ref{fig:fuzzbenchAll},
we additionally provide the code coverage distributions for all fuzzers across all programs within one of the experimental groups. \footnote{The report is available at: \url{https://storage.googleapis.com/www.fuzzbench.com/reports/experimental/2024-05-16-formatfuzz/index.html}.}
\tool achieves the highest performance on five out of 11 programs, while LibAFL
and AFL++ each excel on two programs, and LibFuzzer and FairFuzz excel on one
program each. 

The LLM-generation algorithm demonstrates remarkable effectiveness on certain programs, such as \textit{vorbis_decode_fuzzer}. We illustrate the average code coverage evolution over time for \textit{vorbis_decode_fuzzer} in \F~\ref{fig:fuzzbenchVorbis}. Notably, we observe that \tool achieves higher coverage at 15 minutes compared to all other compared fuzzers at 23 hours. This highlights the capability of \tool to generate diverse and complex structures, enabling the discovery of code regions that are challenging for conventional fuzzers to uncover.


\begin{table}[!htbp]
    \small
    \footnotesize
    \centering
    \caption{Fuzzbench fuzzer ranking. It reports the average rank of fuzzers,
    after we rank them on each benchmark according to their median reached
    code-coverage (lower is better).}
    \label{tab:fuzzbenchRank}
    \vspace{-3pt}
  \setlength{\tabcolsep}{3pt}
    \resizebox{0.75\linewidth}{!}{
    \begin{tabular}{c | c | c | c | c}
    \toprule
    Fuzzers & Group I & Group II & Group III & Average\\
    \midrule
    \tool\ & \textbf{2.09} & \textbf{2.18} & \textbf{2.18}  & \textbf{2.15}\\
    AFL++ & 2.73 & 2.73 & 2.91  & 2.79\\
    LibAFL & 4.55 & 4.55 & 4.64  & 4.58\\
    LibFuzzer & 4.73 & 4.64 & 4.64  & 4.67\\
    HonggFuzz & 6.45 & 6.45 & 6.36  & 6.42\\
    AFLSmart & 6.73 & 6.73 & 6.73  & 6.73\\
    AFL & 7.27 & 7.27 & 7.27  & 7.27\\
    MOPT & 7.27 & 7.27 &  7.27  & 7.27\\
    Eclipser & 7.36 & 7.36 &  7.27 & 7.3\\
    FairFuzz & 8.82 & 8.82 &  8.82 & 8.82\\
    AFLFast & 9.09 & 9.09 &  9.09 & 9.09\\
    Centipede & 9.18 & 9.18 &  9.18 & 9.18\\
    \bottomrule
\end{tabular}
}
\vspace{-10pt}
\end{table} 

\noindent\textbf{Additional Analysis Time.}
Since we run the LLM-generation algorithm before the experiments, \tool has more fuzzing time compared to other fuzzers. To assess its impact, we analyze additional analysis time per program, as shown in \T~\ref{tab:extraTime}. The program with the highest additional time is \textit{bloaty_fuzz_target} at 925 seconds (1.06\% of 23 hours fuzzing time), while \textit{zlib_zlib_uncompress_fuzzer} requires the least at 163 seconds (0.19\%). Eight out of 11 programs need under 500 seconds (0.6\%). We also find that the extra 15 minutes required for LLM-generation had no effect on median code coverage after 23 hours, as shown in \T~\ref{tab:23hVS22h45min}.



\subsubsection{Experiments on MAGMA}
Code coverage and unique bugs are key metrics, but discovering real CVEs directly
shows a fuzzer's ability to find security vulnerabilities with significant 
real-world impact. To avoid bias, we conduct experiments on MAGMA, a ground-truth fuzzing benchmark with real-world bugs for accurate performance evaluation. We integrate \tool into MAGMA and compare it with five fuzzers (i.e., AFL++, MOPT, AFLFast, LibFuzzer, and Entropic). All
AFL++-related fuzzers used in this experiment were based on AFL++ (commit
1d17210) and enabled the RedQueen mutator. 
We excluded programs with textual inputs. \T~\ref{tab:allPrograms} shows the target programs used in MAGMA. Input format types are determined by file name extensions in the initial seeds.
For \textit{openssl}, as MAGMA's initial seeds 
lack extensions, we exclude it from consideration. To avoid randomness, we 
repeat the experiments 5 times.


We analyze the number of CVEs found by each fuzzer, whose results are in
\T~\ref{TTB}. We found that \tool performs the best on the MAGMA benchmark,
uncovering the most bugs in all programs. Specifically, \tool performs the best
on \texttt{libpng_read_fuzzer}, \texttt{tiff_read_rgba_fuzzer}, \texttt{tiffcp},
\texttt{pdf_fuzzer}, \texttt{pdftoppm} and \texttt{pdfimages}, exposing 3, 5, 7,
5, 6 and 6 bugs, respectively. 

\begin{table}[!t]
\centering
\caption{The total CVEs discovered 
(on MAGMA). \tool{} performs the best in discovering real CVEs.}
  \setlength{\tabcolsep}{2.5pt}
    \vspace{-5pt}
\resizebox{0.95\linewidth}{!}{
\begin{tabular}{@{\extracolsep{4pt}}crrrrrrrrrrr@{}}
\toprule
\rotatebox{0}{\textbf{Program}} & \rotatebox{60}{\textbf{\tool{}}}   & \rotatebox{60}{\textbf{AFL++}} & \rotatebox{60}{\textbf{MOPT}} & \rotatebox{60}{\textbf{AFLFast}}  &  \rotatebox{60}{\textbf{LibFuzzer}} & \rotatebox{60}{\textbf{Entropic}} \\
\midrule

libpng_read_fuzzer & \textbf{3} & \textbf{3} & 2 & 1 & \textbf{3} & 0  \\
tiff_read_rgba_fuzzer & \textbf{5} & \textbf{5} & \textbf{5} & 4 & 2 & 3  \\
tiffcp & \textbf{7} & 6 & 4 & 4 & 0 & 0  \\
pdf_fuzzer & \textbf{5} & \textbf{5} & 4 & 2 & 2 & 2  \\
pdftoppm & \textbf{6} & 5 & \textbf{6} & 2 & 0 & 0  \\
pdfimages & \textbf{6} & 5 & 5 & 3 & 0 & 0  \\
\bottomrule

\end{tabular}
}
\label{TTB}
\vspace{-10pt}
\end{table}

\subsubsection{Finding Bugs in Latest Program Versions}

To evaluate \tool's ability to discover real bugs, we test the
latest versions of projects from UNIFUZZ, along with all other executable
programs in these projects suitable for fuzzing. 
Each fuzzer-program pair runs for 24 hours and is repeated 5 times. Following
UNIFUZZ's suggestion, we use the top three functions from the ASAN output to
de-duplicate uncovered bugs. The results are shown in
the~\T~\ref{tab:realBugs}. \tool discovers a total of 10 bugs, while the best
comparative fuzzer, AFL++(cmplog), identifies 5. Notably, 4 bugs are
exclusively discovered by \tool and remain undetected by all other comparative
fuzzers. By the time of writing, we have reported these bugs to the developers, and 3 of them have been confirmed by CVE: \texttt{CVE-2024-57509} (in \texttt{mp42avc}), \texttt{CVE-2024-57510} (in \texttt{mp42avc}), and \texttt{CVE-2024-57513} (in \texttt{mp42hevc}).


\begin{table}[!t]
\centering
\caption{The real bugs discovered by each fuzzer.}
  \setlength{\tabcolsep}{2.5pt}
    \vspace{-5pt}
\resizebox{0.85\linewidth}{!}{
\begin{tabular}{@{\extracolsep{4pt}}crrrrrrrrrrr@{}}
\toprule
\rotatebox{0}{\textbf{Program}} & \rotatebox{60}{\textbf{\tool{}}}   & \rotatebox{60}{\textbf{AFL++(cmplog)}} & \rotatebox{60}{\textbf{AFL++(mopt)}}  &  \rotatebox{60}{\textbf{AFL++(fast)}} & \rotatebox{60}{\textbf{AFL++(rare)}} \\
\midrule

mp3gain          & 1               & 1                       & -                     & 1                     & 1                     \\ 
pdftotext        & 1               & 1                       & 1                     & 1                     & 1                     \\ 
mp42aac          & \textbf{3}               & \textbf{3}                       & 1                     & -                     & -                     \\ 
mp42avc          & \textbf{3}               & -                       & -                     & -                     & 1                     \\ 
mp42hevc         & \textbf{2}               & -                       & -                     & -                     & -                     \\ 
\midrule
Total   & \textbf{10}     & 5              & 2            & 2            & 3            \\ 

\bottomrule

\end{tabular}
}
\label{tab:realBugs}
\vspace{-10pt}
\end{table}


\subsubsection{Classification of Handled File Formats}
\label{sec:HandledFormats}
In the above experiments across three platforms, we use \tool to evaluate its
effectiveness in handling various file formats. As shown in
\T~\ref{tab:formats}, \tool successfully processes a variety of image formats
including JPG, GIF, BMP, and PNG, as well as several audio and video formats
such as MP3, WAV, MP4, and FLV. Additionally, \tool demonstrates capability in
handling PDF documents and various font formats like TTF and OTF. In terms of
file formats, it supports processing formats such as ELF, Mach\_O, and
WebAssembly. All programs associated with the 34 formats were
evaluated in the previous experiments. These findings highlight \tool's strong
performance in fuzz testing across diverse file types. 

The conditions of constructing different file formats vary: some formats are supported
by specific libraries, while others are not. We classify these
conditions into three levels. (1) L1: The target format has specific libraries that can directly generate files.
(2) L2: Some components of the format can be generated using existing libraries,
and these components are then organized
according to the target format's syntax rules.
(3) L3: Files are generated entirely from scratch, based solely on
the target format's syntax rules. Among the 34 tested formats, 23 fall into L1, 3
into L2, and 8 into L3. Formats in L1 typically lead to higher-quality generators
the supporting libraries--often accompanied by documentation, sample code,
and other resources--are included in LLM training data,
increasing both the diversity and accuracy of generated files. Nevertheless, we also
observe decent-quality generators for formats in L2 and L3, which
demonstrates the robustness of \tool in handling various formats.

\begin{table}[h]
    \centering
    \vspace{-5pt}
    \caption{Classification of formats handled by \tool.}
    \label{tab:formats}
    \vspace{-5pt}
    \resizebox{\linewidth}{!}{
    \begin{threeparttable}
    \begin{tabular}{l|lll}
        \toprule
        \textbf{Category} & \textbf{Level} & \textbf{Formats} & \textbf{Related Libraries} \\
        \midrule
        \multirow{12}{*}{Image Formats} & \multirow{8}{*}{L1\tnote{1}} & JPG & PIL/piexif \\
         & & GIF & PIL\\
         & & BMP& PIL\\
         & & PNG& PIL/matplotlib/cv2\\
         & & ICO& PIL\\
         & & XMP& lxml/xml.dom.minidom\\
         & & TGA& PIL\\
         & & TIFF& PIL/tifffile\\
         \cline{2-4}
         & \multirow{3}{*}{L2\tnote{2}} & ANI & PIL\\
         & & RAS& PIL\\
         & & PGX& PIL\\
         \cline{2-4}
         & L3\tnote{3} & PNM/RAW & -\\
        \midrule
        \multirow{8}{*}{Audio Formats} & \multirow{5}{*}{L1} & OGG & soundfile\\
        &  & MP3 & pydub/mutagen\\
        &  & WAV & wave/scipy\\
        &  & AIFF & soundfile/pydub/wave\\
        &  & AIFC & aifc\\
        \cline{2-4}
         & L3 & AU/CAF & -\\
        \midrule
        \multirow{2}{*}{Video Formats} & \multirow{2}{*}{L1} &FLV & moviepy\\
        &  & MP4 & cv2/moviepy/mutagen\\
        \midrule
        Document Formats & L1 & PDF & fpdf/PyPDF2/reportlab\\
        \midrule
        
        Font Formats & L1 & TTF/OTF/WOFF/TTC & fontTools\\
        \midrule
        \multirow{4}{*}{File Formats} & \multirow{3}{*}{L1} & Zlib compressed & zlib\\
        &  & PCAP & scapy\\
        &  & DER certificate & cryptography\\
        \cline{2-4}
         & L3 & ELF/Mach O/WebAssembly/ICC profile & -\\
        \bottomrule
    \end{tabular}
    \begin{tablenotes}
    \item[1] \textbf{L1}: Use specialized libraries to create files in the target format.
    \item[2] \textbf{L2}: Construct parts of the file with specific libraries and organize them according to the target format's syntax rules.
    \item[3] \textbf{L3}: Build the file from scratch based on the target format's rules, directly writing binary data or using \textit{struct} to write the data.
    \end{tablenotes}
    \end{threeparttable}
    }
    \vspace{-10pt}
\end{table}


\subsection{Compared with Structure-Aware Fuzzers}

We compare \tool with the SOTA grammar-aware fuzzer FormatFuzzer and the SOTA
inference-based fuzzer WEIZZ using the UNIFUZZ benchmark. We do not compare
against AFLSmart because it has already been compared in FuzzBench, where \tool
significantly outperforms AFLSmart. \tool's average code coverage rank is 2.15,
while AFLSmart's is 6.73. Due to the considerable gap, we do not conduct
additional experiments here. We do not compare against Superion, Nautilus, and
Grimoire, as these fuzzers have only been assessed on text-based grammar input
formats. Moreover, we do not include FuzzInMem, ProFuzzer, and GreyOne because
they have not been made open source.


\begin{table}[!htbp]
    \small
    \footnotesize
    \centering
    \caption{The average line coverage discovered by \tool, FormatFuzzer, and WEIZZ.}
    \vspace{-5pt}
    \label{tab:structure}
    \resizebox{\linewidth}{!}{
    \begin{threeparttable}
    \begin{tabular}{c | c | c | c | c | c | c }
    \toprule
    \multirow{2}{*}{Programs} & \multicolumn{2}{c|}{\tool} & \multicolumn{2}{c|}{FormatFuzzer} & \multicolumn{2}{c}{WEIZZ} \\
    & line & function & line & function & line & function \\
    \midrule
    exiv2& \textbf{5,984} & \textbf{1488} & 1,138 & 369 & 3,732 & 1025\\
    ffmpeg& \textbf{53,664} & \textbf{3028} & 23,114 & 1554 & 26,789 & 1795\\
    flvmeta& 623 & 59 & -\tnote{1} & -& \textbf{632} & \textbf{60}\\
    imginfo& \textbf{5,003} & \textbf{364} & 2,128 & 193 & 3,481 & 275\\
    jhead& \textbf{431} & \textbf{21} & 239 & 16 & 300 & 18\\
    mp3gain& \textbf{2,168} & \textbf{58} & \#\tnote{2} & \# & 2,103 & 56\\
    mp42aac& \textbf{3,378} & \textbf{811} & \# & \# & 2,041 & 504\\
    pdftotext& \textbf{13,733} & \textbf{1182} & - & - & 9,133 & 914\\
    tiffsplit& \textbf{3,176} & \textbf{194} & - & - & 3,019 & 185\\
    gdk& \textbf{4,856} & \textbf{315} & 2,287 & 192 & =\tnote{3} & =\\
    \bottomrule
    \end{tabular}
    \begin{tablenotes}
    \item[1] \textbf{-}: FormatFuzzer does not support PDF, TIFF, and FLV formats.
    \item[2] \textbf{\#}: We encountered issues while running FormatFuzz.
    \item[3] \textbf{=}: We are unable to compile \textit{gdk-pixbuf} by WEIZZ.
    \end{tablenotes}
    \end{threeparttable}
}
\vspace{-5pt}
\end{table} 

\begin{table}[!htbp]
    \small
    \footnotesize
    \centering
    \vspace{-5pt}
    \caption{Functions exclusively discovered by each fuzzer.}
    \vspace{-5pt}
    \label{tab:uniqueFunc}
  \setlength{\tabcolsep}{3pt}
    \resizebox{0.6\linewidth}{!}{
    \begin{tabular}{c | c | c | c }
    \toprule
    Program & \tool & FormatFuzzer & WEIZZ\\
    \midrule
    gdk & \textbf{126} & 0 & - \\
    exiv2 & \textbf{424} & 0 & 0 \\
    ffmpeg & \textbf{1951} & 6 & 14 \\
    flvmeta & 0 & - & \textbf{1} \\
    imginfo & \textbf{148} & 0 & 0 \\
    jhead & \textbf{8} & 0 & 0 \\
    mp3gain & \textbf{1} & - & 0 \\
    mp42aac & \textbf{338} & - & 0 \\
    pdftotext & \textbf{334} & - & 0 \\
    tiffsplit & \textbf{10} & - & 0 \\
    \bottomrule
\end{tabular}
}
\vspace{-10pt}
\end{table}

As \tool, FormatFuzzer, and WEIZZ use different instrumentation methods, they may achieve varying edge coverage levels with identical inputs. To
accurately measure line coverage, we utilize \textit{afl-cov}. The results are
presented in \T~\ref{tab:structure}.
To clarify, we encountered issues running some programs with FormatFuzzer and
WEIZZ. Specifically, FormatFuzzer generated an excessive number of
\textit{core.*} files while testing \textit{mp42aac}, consuming over 500GB of
memory within 10 hours. Additionally, we face errors when building MP3
generators for mp3gain according to FormatFuzzer's instructions. FormatFuzzer is
also unsuitable for testing \textit{pdftotext}, \textit{tiffsplit}, and
\textit{flvmeta}, as it does not support PDF, TIFF, and FLV formats. As for
WEIZZ, we are unable to compile \textit{gdk-pixbuf}. 

For nine out of 10 programs, \tool achieves higher line coverage than both FormatFuzzer and WEIZZ. 
For example, \tool achieves more than twice as much line coverage as FormatFuzzer in \textit{exiv2}, \textit{ffmpeg}, \textit{imginfo}, \textit{gdk}.
Unlike the grammar-based fuzzer FormatFuzzer, \tool is scalable to a broader range of programs that accept different formats. Moreover, it is common for a program to accept multiple input formats, but FormatFuzzer can handle only one format at a time, which can reduce the diversity of generated inputs.

To further validate that the files generated by \tool with complex features
help uncover more intricate program logic, we measure the number of functions
exclusively discovered by each fuzzer. Here, ``exclusive'' refers to functions
that are not detected by any other fuzzer. A fuzzer that finds more exclusive
functions illustrates its ability to trigger more subtle program logics. The
results are shown in the \T~\ref{tab:uniqueFunc}. For nine out of ten programs,
\tool identifies the largest number of exclusive functions, confirming the
effectiveness of using LLMs to generate complex binary inputs.

    


\subsection{Compared with Fuzztruction}

To compare with Fuzztruction, we use \tool to generate a batch of initial seeds
and conduct experiments in the Docker environment provided by Fuzztruction,
ensuring that all experimental parameters remain consistent. We test nine programs
used by Fuzztruction; however, three lack clear input file extensions and are thus
incompatible with \tool. The tests run for 6 hours and are repeated 5 times, and the versions of all test programs are consistent with the versions tested by Fuzztruction. The mean coverage is shown
in~\T~\ref{tab:FuzztructionEdgeBug}. Overall, \tool outperforms Fuzztruction on
seven out of nine programs. \tool semantically constructs files with different
structures from scratch, while Fuzztruction's generator--primarily a
converter--still requires structured initial seeds, and its bit-level mutation
only makes subtle adjustments. However, Fuzztruction performs better on programs using zip files. We believe \tool finds fewer bugs due to the limited functionality of Python libraries for constructing zip files, restricting coverage of file characteristics.

We compare the seed quality of \tool and Fuzztruction by measuring feature coverage. In this experiment, \tool generates seeds only during the initial stage, while Fuzztruction continuously generates seeds. 
To ensure fairness, we compare the feature coverage of \tool and Fuzztruction using the same number of generated seeds, with the number of seeds generated by \tool serving as the baseline. For program selection, we target all programs that take image or document inputs, including pngtopng, pdftotext, and qpdf. In Fuzztruction, PDFs generated by the qpdf generator are excluded due to unparseable password options. The results are shown in the~\T~\ref{tab:FuzztructionFeatureCov}, revealing that
\tool discovers more unique features than Fuzztruction.
In terms of validity ratio, \tool achieves a higher validity ratio for PNGs and PDFs compared to Fuzztruction.
Additionally, \tool discovers more rare features, such as \textit{Properties-Digital Signature} and \textit{Properties-png:PLTE.number_colors} in PNG files, which Fuzztruction cannot cover.

\begin{table}[!htbp]
    \small
    \footnotesize
    \centering
    \vspace{-5pt}
    \caption{The average coverage (in basic blocks) and bugs discovered by Fuzztruction
    and \tool. }
    \vspace{-5pt}
    \label{tab:FuzztructionEdgeBug}
  \setlength{\tabcolsep}{3pt}
    \resizebox{0.9\linewidth}{!}{
    \begin{tabular}{c | c | c | c | c | c}
    \toprule
    \multirow{2}{*}{Input Format} & \multirow{2}{*}{Program} & \multicolumn{2}{c|}{Fuzztruction} & \multicolumn{2}{c}{\tool}\\
    &&Cov & Bug & Cov & Bug\\
    \midrule
    pdf                   & pdftotext-enc    & 36853.8 & 0           & \textbf{38866.0} & 0 \\
pdf                   & pdftotext        & 35108.4 & 0           & \textbf{39011.4} & 0 \\
elf                   & objdump          & 12468.8 & 0           & \textbf{12851.8} & 0 \\
elf                   & readelf          & 12347.8 & 0           & \textbf{13328.2} & 0 \\
png                   & pngtopng         & 4414.6 & 0            & \textbf{4566.2} & 0  \\
der                   & vfychain         & \textbf{14937.4} & 0  & 11600.4 & 0          \\
7z                    & 7zip-enc         & 28887.2 & \textbf{8}           & \textbf{28909.0} & 6 \\
zip                   & 7zip             & \textbf{34585.4} & \textbf{8}  & 31691.4 & 6          \\
zip                   & unzip            & 2788.2 & 1            & \textbf{3104.8} & 1  \\
\bottomrule
\end{tabular}
}
\vspace{-10pt}
\end{table} 

\begin{table}[!htbp]
    \small
    \footnotesize
    \centering
    \vspace{-5pt}
    \caption{Functions exclusively discovered by each fuzzer.}
    \vspace{-5pt}
    \label{tab:FuzztructionFeatureCov}
  \setlength{\tabcolsep}{3pt}
    \resizebox{\linewidth}{!}{
    \begin{tabular}{c | c | c | c |c}
    \toprule
     & \multicolumn{2}{c|}{PNG(pngtopng)} & \multicolumn{2}{c}{PDF(pdftotext)}\\
     & Feature Cov & ValidNum/InvalidNum & Feature Cov & ValidNum/InvalidNum\\
    \midrule
    \tool & \textbf{457.0} & \textbf{36/0} & \textbf{531.0} & \textbf{50/12}\\
    Fuzztruction & 427.4 & 27.8/8.2 & 152.4 & 48.8/13.2\\
    
    \bottomrule
\end{tabular}
}
\vspace{-10pt}
\end{table} 

\subsection{Feature Coverage}
To verify whether \tool can generate file features that other fuzzers cannot cover, we compare the feature coverage of each fuzzer. The calculation of feature coverage is challenging due to the lack of a unified quantification method. Therefore, we use \textit{ImageMagick} to extract each seed's attributes, such as compression type, treating each attribute as a feature. We then manually remove irrelevant attributes that vary across most files, such as the file name.

We select four formats--TIFF, JPG, MP4, and PDF--from Sec.~\ref{sec:unifuzz} for analysis,
covering image files, video files, and complex documents. The results are shown
in the~\T~\ref{tab:featureCov}. \tool (using GPT-4) achieves the highest feature coverage for
TIFF, JPG, and MP4. AFL++ focuses more on low-level mutations, which struggle to
modify high-level features. Changing high-level features requires handling the
constraints across multiple chunks simultaneously, which is highly challenging
for byte-level mutation. In contrast, \tool can semantically mutate the
generator or generate seeds with the target features from scratch, enabling
broader coverage of high-level file features. Note that
\textit{ImageMagick} can only parse valid inputs, while most seeds generated by
AFL++ mutations are invalid, resulting in lower feature coverage being recorded
for AFL++. For example, the mutations of AFL++ fail to produce valid MP4 files,
resulting in a feature coverage of 0. 
In the PDF format (\textit{pdftotext}), AFL++ (rare) and AFL++ (fast) cover more features in terms of skewness, kurtosis, and standard deviation in the Blue/Green/Red channels. Seeds with such features receive higher weights in AFL++ (rare) and AFL++ (fast), leading to more frequent mutations and, consequently, higher coverage of these features. However, from the perspective of final code coverage, allocating excessive energy to explore such features is inefficient.

\tool can construct some rare features, such as
\textit{Properties-tiff:timestamp}, \textit{Properties-tiff:copyright}, and
\textit{Properties-Contact} in TIFF files. Furthermore, for
\textit{Chromaticity-Compression}, \tool can cover all four compression methods--\textit{Zip}, \textit{RLE}, \textit{JPEG}, and \textit{LZW}--whereas other
fuzzers can only cover \textit{RLE} and \textit{JPEG}. We observe that
covering rare features can better trigger specific program logic in the
target program, thereby improving code coverage.

\begin{table}[!t]
\centering
\caption{The feature coverage covered by each fuzzer.}
  \setlength{\tabcolsep}{2.5pt}
    \vspace{-5pt}
\resizebox{0.85\linewidth}{!}{
\begin{tabular}{@{\extracolsep{4pt}}crrrrrrrrrrr@{}}
\toprule
\rotatebox{0}{Program} & \rotatebox{60}{\textbf{\tool{}}}   & \rotatebox{60}{\textbf{AFL++(cmplog)}} & \rotatebox{60}{\textbf{AFL++(mopt)}}  &  \rotatebox{60}{\textbf{AFL++(fast)}} & \rotatebox{60}{\textbf{AFL++(rare)}} \\
\midrule

TIFF (tiffsplit)          & \textbf{4719}                    & 2303                    & 2169                  & 2369                  & 2178                  \\ 
JPG (exiv2)               & \textbf{4096}                    & 1910                    & 1910                  & 1913                  & 1909                  \\ 
MP4 (mp42aac)             & \textbf{1459}                    & 0                       & 0                     & 0                     & 0                     \\ 
PDF (pdftotext)           & 37797                   & 33451                   & 40730                 & \textbf{44704}                 & 36203                 \\ 

\bottomrule

\end{tabular}
}
\label{tab:featureCov}
\vspace{-10pt}
\end{table}

\subsection{Generalizability Across LLMs}
To demonstrate \tool's generalizability across different LLMs, we select the open-source models \textit{llama-3-8b-instruct} and \textit{llama-3-70b-instruct} for our experiments. Under the same setup, these models generate initial seeds for five file formats, completing \textit{Input Generator Synthesis} and \textit{Generator Mutation} during the initialization stage. 
For \textit{GPT-3.5} and \textit{GPT-4}, we reuse the initial seeds generated from the first epoch of experiments for JPG, TIFF, MP3, MP4, and PDF with \tool on \textit{exiv2}, \textit{tiffsplit}, \textit{mp3gain}, \textit{mp42aac}, and \textit{pdftotext}. Only files with the target format suffix are considered, as the generator may produce files in other formats.

The results are shown in the~\T~\ref{tab:diffLLMs}. \textit{GPT-4} achieves the
highest feature coverage for JPG and PDF, while the open-source models
\textit{llama-3-8b-instruct} and \textit{llama-3-70b-instruct} achieve the
highest feature coverage for TIFF and MP4, respectively. Notably,
\textit{llama-3-70b-instruct} outperforms \textit{GPT-3.5} across all four formats.
These results demonstrate \tool's scalability and its ability to
generate high-quality generators using open-source models. 

We also evaluate the effectiveness of the prompt used by \tool with 10 different
formats (including images, videos, and documents) and find that GPT-4 performs
well across most formats. More details can be found in
Appendix~\ref{sec:PromptEffectiveness}. Additionally, we analyze the impact of
different libraries on the generator quality and find that collaboration between
multiple libraries is the most efficient approach. For more details, please
refer to Appendix~\ref{sec:LibraryInfluence}. 

\begin{table}[!htbp]
    \small
    \footnotesize
    \centering
    \vspace{-5pt}
    \caption{Functions exclusively discovered by each fuzzer.}
    \vspace{-5pt}
    \label{tab:diffLLMs}
  \setlength{\tabcolsep}{3pt}
    \resizebox{0.9\linewidth}{!}{
    \begin{threeparttable}
    \begin{tabular}{c | c | c | c | c}
    \toprule    
    Format & GPT-3.5 & GPT-4 & llama-3-8b-instruct & llama-3-70b-instruct \\
\midrule
JPG  & 259 & \textbf{984} & 211 & 636 \\
MP4  & -\tnote{1}   & 290 & 245 & \textbf{517} \\
PDF  & 374 & \textbf{559} & 555 & 504 \\
TIFF & 388 & 387 & \textbf{591} & 516 \\
    \bottomrule
\end{tabular}
\begin{tablenotes}
\item[1] \textbf{-}: \textit{GPT-3.5} cannot generate valid MP4 files during the first round due to randomness.
\end{tablenotes}
\end{threeparttable}
}
\vspace{-15pt}
\end{table}

\section{Discussion}


\tool supports only file formats that have accompanied
generator libraries available. Nevertheless, it can integrate with user-written
file format specifications (by using prompts such as ``\textit{generate Python
generator code based on the provided format specifications.}''). Thus,
supporting corner cases or new file formats requires only extra engineering and
manual effort. More importantly, we see an encouraging trend of emerging Python
libraries for file generation (searching for JPEG and MP4 on GitHub reveals 21
and 26 file generation/editing libraries, respectively, created within the past
three years); this illustrates the high extensibility of \tool to adapt to new
formats without code changes. Overall, leveraging existing and emerging
libraries, \tool can support more file formats, thus improving fuzzing
efficiency in a broader range of scenarios and continuous manner. 
In addition, \tool is currently unable to handle custom formats. This limitation
could be alleviated by adding document parsing capabilities, allowing LLMs to
learn and adapt to custom syntax; we leave this as future work.





\section{Conclusion}

In this paper, we present \tool, a novel and highly efficient approach that
augments mutation-based fuzzing with LLMs. We identify a unique opportunity to combine
the strengths of LLMs and mutation-based fuzzers to achieve a synergistic effect.
The evaluation shows that \tool consistently outperforms SOTA mutation-based fuzzers
and several other fuzzer baselines.
\section{Acknowledgment}
We sincerely thank the anonymous reviewers and our shepherd for their valuable feedback and guidance. The HKUST authors are supported in part by a RGC GRF grant under the contract 16214723. 
\section{Ethics Considerations}
\noindent\textbf{Vulnerability Disclosure.}
Our fuzzing tool is designed to uncover security vulnerabilities in software. If we identify such vulnerabilities, failing to disclose them responsibly could lead to serious security risks, such as unauthorized access or exploitation by malicious actors.

We have established a responsible disclosure process. When we find vulnerabilities, we promptly notify the affected software vendors, giving them sufficient time to patch the issues before making any public disclosures. This ensures our research contributes positively to security without exposing users to unnecessary risks.


\noindent\textbf{Experiments with Live Systems Without Informed Consent.}
If we apply our fuzzing tool to live systems or real-world software without obtaining consent from the owners or operators, this could disrupt services or negatively impact users who rely on those systems.

We avoid testing live systems without explicit consent from the system owners. When testing on live systems is necessary, we first obtain informed consent and design our testing methods to minimize any potential harm or disruption. This approach respects the rights and interests of those who rely on the systems we study.

\noindent\textbf{Terms of Service.}
Our tool could potentially violate the terms of service of the software we are testing, particularly if the software explicitly prohibits automated testing or fuzzing. This could lead to legal issues or harm our reputation in the research community.

Before conducting any fuzzing, we thoroughly review the terms of service of the software being tested. If our activities might violate these terms, we seek permission from the software provider or adjust our methods to avoid violations. This ensures our research is both ethical and legally compliant.

\noindent\textbf{Deception.}
If our testing involves any form of deception, such as obscuring the true nature of the tests from the system administrators or users, this could raise ethical concerns, particularly if it results in harm or a loss of trust.

We avoid using deception in our research. If deception is necessary for the validity of the study, it is ethically justified and followed by a thorough debriefing to explain the research's purpose and methods to all affected parties. This approach maintains transparency and trust.

\noindent\textbf{Wellbeing for Team Members.}
Our work may expose team members to stressful or disturbing content, especially when analyzing malicious software, which could impact their psychological wellbeing.

We prioritize our team’s wellbeing by supporting those exposed to stressful content, setting clear boundaries, and maintaining a safe, supportive work environment.

\noindent\textbf{Innovations with Both Positive and Negative Potential Outcomes.}
The tools and techniques we develop have potential for misuse by adversaries. While our intention is to improve software security, there is a risk that others could use our tool to find and exploit vulnerabilities for malicious purposes.

We recognize the dual-use nature of our tool and have implemented safeguards to prevent misuse. Access is restricted, and we work with ethical review boards to assess risks. We also engage with the security community to ensure our research is used positively.

\noindent\textbf{Retroactively Identifying Negative Outcomes.}
If our research unintentionally causes negative outcomes, like service disruptions or exploitation of vulnerabilities, failing to address them could harm users and damage our credibility.

We will monitor for any issues and take responsibility if they arise, working to remediate any harm. This proactive approach ensures our research remains ethical and responsible.

\noindent\textbf{The Law.}
Our fuzzing activities must comply with cybersecurity laws and regulations. Any inadvertent violations could lead to legal consequences for us and our institution.

We consult legal experts to ensure compliance and obtain necessary approvals before engaging in risky activities, minimizing legal risks and ensuring proper conduct.

\section{Open Science}


To ensure compliance with open science principles, we commit to making our research data, code, and materials publicly accessible through publicly available repositories. This includes providing access to our tool code and experiment data at \url{https://github.com/G2FUZZ/G2FUZZ}\footnote{Our tool code is also available at \url{https://zenodo.org/records/14728879}.} and \url{https://github.com/G2FUZZ/G2FUZZ-DATA}, allowing others to review, utilize, and adapt our implementation.

We also document our research methods, experiments, and results in detail to enable reproducibility. All relevant information will be shared openly to allow other researchers to replicate and build upon our work.

\bibliographystyle{plain}
\bibliography{bib/ref,bib/similarity,bib/decompiler,bib/machine-learning,bib/attack}

\appendix

\section{Challenges in Generating Complex Files: A TIFF Case Study}
\label{sec:caseStudy}
\begin{figure}[h] 
    \centering
    \begin{minipage}[t]{0.49\textwidth}
        \centering
        \includegraphics[width=\textwidth]{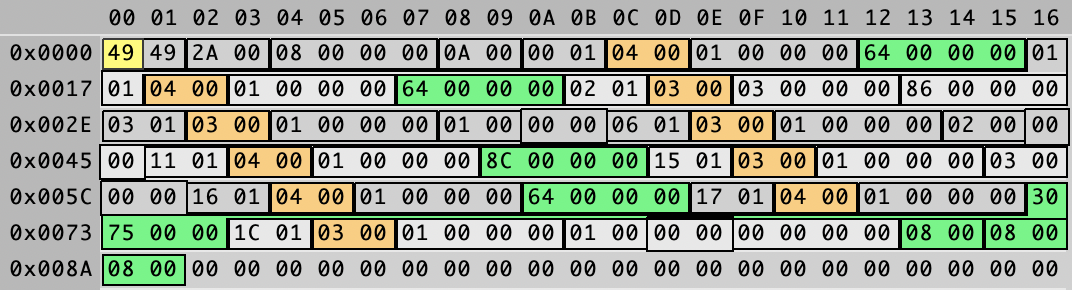}
        \vspace{-15pt}
        \subcaption{Original TIFF file.}
        \label{fig:seedCompress1}
    \end{minipage}
    \begin{minipage}[t]{0.49\textwidth}
        \centering
        \includegraphics[width=\textwidth]{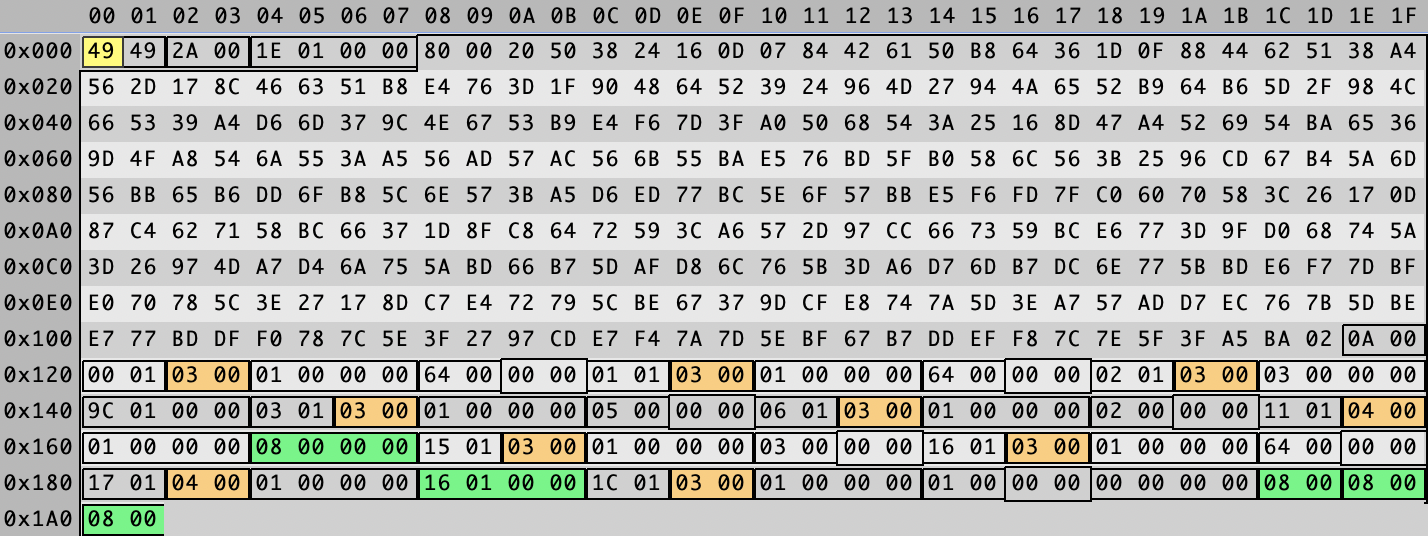}
        \vspace{-15pt}
        \subcaption{TIFF file with LZW compression enabled.}
        \label{fig:seedCompress2}
    \end{minipage}
    \vspace{-10pt}
    \caption{Comparing two TIFF files with LZW compression enabled or not, both
    containing an identical image data. The newly added LZW-related chunks in
    \F~\ref{fig:seedCompress2} from (0x000, 0x08) to (0x100, 0x1D) cannot be
    parsed without specifications.}
    \vspace{-25pt}
    \label{fig:seedCompress}
\end{figure}

\begin{figure}[h] 
    \centering
    \includegraphics[width=0.35\textwidth]{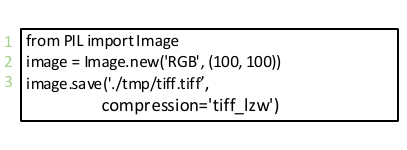}
    \vspace{-20pt}
    \caption{Python generator for creating the TIFF file with LZW compression data in \F~\ref{fig:seedCompress2}.}
    \vspace{-10pt}
    \label{fig:tiffCompressCode}
\end{figure}
In~\S~\ref{sec:motivation}, we argue that generating files with complex features is challenging for current fuzzers. To illustrate this, we provide an example.
TIFF, which stands for Tagged Image File Format, is a flexible and adaptable
file format for storing images. Note that TIFF files support various compression
algorithms. Here, we analyze the use of LZW compressed data within TIFF files to
clarify why generating files with complex features is hard for existing fuzzers.
\F~\ref{fig:seedCompress} illustrates the differences between two TIFF files
with an \textit{identical} image data: \F~\ref{fig:seedCompress1} shows the
original TIFF file, whereas \F~\ref{fig:seedCompress2} shows the file with LZW
compression enabled. Two main differences exist: \textit{1. introducing many
(unparsed) data blocks.} In \F~\ref{fig:seedCompress2}, a large data block is
introduced. Note that it is ``unparsed'' because the LZW specification is
missing in the 010 Editor template used by FormatFuzzer, which prevents parsing
and further mutating. \textit{2. changes in data values and new constraints:}
Many data values in \F~\ref{fig:seedCompress2} have changed, with these changes
introduced new constraints (e.g., offsets and sizes) that need to be met. For
example, when adding Exif features to a TIFF file, an ExifIFDPointer tag is
added to the primary IFD to refer the Exif data. The Exif data, like color
space, must be aligned with those in TIFF data, introducing new constraints
between the Exif data and the primary IFD.

Based on our exploration, current binary-format fuzzers cannot generate
TIFF files containing LZW data. These methods can be categorized into two types:
\textit{1. inference-based fuzzing}, such as WEIZZ. WEIZZ infers input fields
and an approximate structure of the chunks on-the-fly while mutating. The
inference results can be inaccurate, failing to precisely capture relations
between chunks and making it unsuitable for constructing files with complex
features. \textit{2. grammar-aware fuzzing}, such as FormatFuzzer and AFLSmart.
They rely on user-provided grammars to parse and mutate inputs. However, the
standard TIFF specification can be insufficient frequently, and specifications
of other formats are needed. In particular, due to the absence of LZW syntax in
the grammar files shipped by FormatFuzzer and AFLSmart, they cannot generate
TIFF files that include LZW data. Thus, even if an initial TIFF seed contains
compression data, existing methods still cannot parse and mutate it.

Overall, complex features are very common across various file formats in
different domains, such as the complex Exif data in JPEG files,
transparency capabilities in PNGs, and encryption and DRM protection in MP4
files. It's worth noting that these complex features often involve more
intricate logic and state management, which may likely result in security
vulnerabilities. Therefore, constructing test input files with various complex
features is crucial for enhancing fuzzing.

\begin{table}[h]
    \small
    \footnotesize
    \centering
    \caption{Token consumption and cost analysis for 24 hours of fuzzing in UNIFUZZ.}
    \label{tab:token}
    \vspace{-5pt}
    \resizebox{0.9\linewidth}{!}{
    \begin{tabular}{c | c | c | c | c}
    \toprule
    \multirow{2}{*}{Programs} & \multicolumn{2}{c|}{\textsc{G$^2$Fuzz}(GPT-3.5)} & \multicolumn{2}{c}{\textsc{G$^2$Fuzz}(GPT-4)} \\
    & Token Count & Cost(\$) & Token Count & Cost(\$)\\
    \midrule
    ffmpeg & 57,870.0 & 0.10 & 112,763.8 & 3.97\\
    gdk & 80,720.4 & 0.14 & 186,810.8 & 6.50\\
    jhead & 68,816.2 & 0.12 & 158,385.6 & 5.68\\
    mp42aac & 64,593.6 & 0.12 & 170,129.4 & 6.03\\
    tiffsplit & 75,616.6 & 0.14 & 156,247.8 & 5.58\\
    exiv2 & 41,958.4 & 0.08 & 100,312.0 & 3.58\\
    flvmeta & 52,277.4 & 0.09 & 350,607.6 & 12.71\\
    imginfo & 66,237.4 & 0.12 & 138,265.4 & 4.78\\
    mp3gain & 108,270.2 & 0.20 & 152,863.2 & 5.50\\
    pdftotext & 69,787.6 & 0.13 & 131,047.2 & 4.64\\
    \bottomrule
\end{tabular}
}
\vspace{-10pt}
\end{table} 

\section{Token and Cost Analysis}
\label{sec:TokenCost}

We further evaluate the token cost of LLMs for fuzzing. Overall, as we do not
rely on LLMs to perform mutation engines, \tool does not need too many tokens.
We collect the token consumption of GPT-3.5 and GPT-4 in the experiments of
UNIFUZZ. The results are shown in \T~\ref{tab:token}. In all programs,
\textsc{G$^2$Fuzz}(GPT-3.5) costs less than 0.2\$ for a 24 hour fuzzing process,
while \textsc{G$^2$Fuzz}(GPT-4) costs less than 13\$. We interpret that the
token cost is acceptable for fuzzing.

\section{Ablation Study}
\label{sec:Ablation}
\subsection{The contribution of each component}

As \tool comprises two main components: input generator synthesis and generator
mutation, we analyze the contribution of each component. Our goal is to assess
the effectiveness of the seeds generated by these components. If mutating a seed
results in the discovery of a new path, we consider it useful. Therefore, we
count the number of new paths found by mutating seeds from each component. The
component that contributes more new paths is deemed more effective.

The results are presented in \T~\ref{tab:ablation}. On average, both the input
generator synthesis and the generator mutation have proven to be effective. In
total, the input generator synthesis contributes 82,001 new paths, while the
generator mutation contributes 141,340. Specifically, in \textit{jhead}, the
input generator synthesis is responsible for discovering almost all the new
paths. In \textit{tiffsplit}, \textit{ffmpeg}, \textit{exiv2}, and
\textit{mp3gain}, the generator mutation contributes the most to discovering new
paths.

\begin{table}[!htbp]
    \small
    \footnotesize
    \centering
    \caption{The number of new pathes contributed by the different components of
    \tool.}
    \vspace{-5pt}
    \label{tab:ablation}
    \resizebox{0.9\linewidth}{!}{
    \begin{tabular}{c | c | c | c }
    \toprule
    \multirow{2}{*}{Programs} & \multirow{2}{*}{Initial Seeds} & Input Generator  & \multirow{2}{*}{Generator Mutation}  \\
    &&Synthesis&\\
    \midrule
    tiffsplit & 101 & 539 & 2,549\\
    jhead & 2 & 1,046 & 0 \\
    mp42aac & 6,859 & 522 & 6,298 \\
    gdk & 9,832 & 12,993 & 1\\
    ffmpeg & 4,877 & 14,603 & 109,381\\
    exiv2 & 398 & 40,719 & 18,328\\
    flvmeta & 1,133 & 1,066 & 29\\
    imginfo & 21,236 & 593 & 0\\
    mp3gain & 1,675 & 1,377 & 4,266\\
    pdftotext & 7,735 & 8,543 & 488\\
    \midrule
    Total & 53,848 & 82,001 & 141,340\\
    \bottomrule
\end{tabular}
}
\vspace{-10pt}
\end{table}

\subsection{Compared with LLM-Only \tool}

In \tool, we leverage LLMs for generating diverse seeds and performing mutations using traditional byte-level techniques. Previous experiments confirm the LLM’s effectiveness in seed generation. To assess the need for combining LLMs with traditional methods, we created \textsc{G$^2$Fuzz}(LLM-Only), which solely relies on LLMs for seed mutation. Testing on UNIFUZZ shows that \textsc{G$^2$Fuzz}(LLM-Only) finds fewer edges and has lower throughput, often less than 1\% of \tool, as shown in \T~\ref{tab:llmOnly}. It also struggles with low-level mutations and is significantly more expensive, making the integration of LLMs and traditional fuzzing both necessary and efficient.

\begin{table}[!htbp]
    \small
    \footnotesize
    \centering
    \caption{The evaluation of \textsc{G$^2$Fuzz}(LLM-Only).}
    \label{tab:llmOnly}
    \resizebox{0.9\linewidth}{!}{
    \begin{tabular}{c | c | c | c | c }
    \toprule
    \multirow{2}{*}{Programs} & \tool & \multirow{2}{*}{Throughput} & \multirow{2}{*}{Token Count} & \multirow{2}{*}{Cost(\$)} \\
    &(LLM-Only)&&&\\
    \midrule
    flvmeta & 150 & 5,758,008 & 16,253,249 & 15.72\\
    exiv2 & 2,126 & 7,643 & 20,831,178 & 17.80\\
    gdk & 830 & 18,007 & 18,253,243 & 16.63\\
    imginfo & 909 & 13,155 & 17,353,735 & 16.10\\
    jhead & 245 & 150,634 & 18,523,648 & 17.00\\
    mp42aac & 728 & 11,645 & 16,349,895 & 15.13\\
    tiffsplit & 938 & 6,729 & 20,539,769 & 17.75\\
    mp3gain & 691 & 9,337 & 18,547,650 & 16.31\\
    pdftotext & 3,300 & 3,084 & 22,996,919 & 19.20\\
    \bottomrule
\end{tabular}
}
\vspace{-5pt}
\end{table}

\section{Prompt Effectiveness}
\label{sec:PromptEffectiveness}
To evaluate the effectiveness of the prompts we used, we analyze
three attributes. 1) Validity: The generator produced by \tool should 
be able to construct valid seeds. 2) Proportion of Seeds with the Target Feature
(PSTF): Seeds that contain the necessary code to produce the target feature are
deemed to possess it. 3) Proportion of Unique, Useful Features (PUUF).

We analyze all generators from Sec.~\ref{sec:unifuzz} for validity and manually
review the generators produced during \textit{Input Generator Synthesis} for
the other two attributes. Specifically, we exclude files whose suffix matches the target format
and use \texttt{ImageMagick} for analysis. Note that
\texttt{ImageMagick} can process various image formats as well as PDF and MP4
(see \T~\ref{tab:promptEff}). For example, for TIFF, we exclude seeds with a
TIFF suffix from all programs, then parse each one with \textit{ImageMagick}.
Seeds that can be parsed are deemed valid, and vice versa. The results are shown
in~\T~\ref{tab:promptEff}. GPT-4 achieves a validity rate exceeding 80\% across all 10
formats, with PSTF over 70\% in 5 formats and PUUF above 70\% in 8
formats. These findings demonstrate the effectiveness of \tool's prompts in
efficiently accomplishing the target tasks.

The unsuccessful outcomes can be attributed to four reasons: 1) LLM
hallucinations generate non-existent features. 2) Debugging (Alg.
\ref{alg:programGeneration}) leads the LLM to remove code related to the target
feature for proper execution. 3) Rare features are harder to generate. 4) Some
features exist in all files of a given type, rendering them useless. We further
analyze the impact of LLM hallucinations on \tool. During the feature generation
phase, hallucinations are relatively rare, with most useless features like
``default features describing the target format'' or ``redundant features.''
Hallucinations primarily occur during generator synthesis, often referencing
non-existent functions or attributes and triggering exceptions such
as \textit{AttributeError} or \textit{NotImplementedError}. However, due to our
debugging strategy (Alg. \ref{alg:programGeneration}), these errors caused by hallucinations are promptly detected
when executing the generator. The LLM then attempts to fix them,
effectively mitigating the impact of hallucinations.

\begin{table}[t]
    \small
    \footnotesize
    \centering
    \caption{Analysis of prompt effectiveness for different formats.}
    \label{tab:promptEff}
  \setlength{\tabcolsep}{3pt}
    \resizebox{0.9\linewidth}{!}{
    \begin{tabular}{c | c | c | c | c}
    \toprule    
    \multirow{2}{*}{Format} & \multicolumn{2}{c|}{Validity Rate} & \multirow{2}{*}{PSTF (GPT-4)} & \multirow{2}{*}{PUUF (GPT-4)}\\
    & GPT-3.5 & GPT-4 & & \\
\midrule
TIFF & 91.90\% & 94.31\% & 90.00\% & 90.00\% \\
BMP & 57.25\% & 97.26\% & 50.00\% & 60.00\% \\
JPG & 98.05\% & 99.16\% & 70.00\% & 80.00\% \\
PNG & 90.78\% & 99.65\% & 77.77\% & 77.77\% \\
GIF & 88.37\% & 100\% & 75.00\% & 75.00\% \\
ICO & 47.88\% & 100\% & 57.14\% & 85.71\% \\
TGA & 22.22\% & 82.81\% & 88.88\% & 88.88\% \\
PNM & 67.12\% & 90.00\% & 37.50\% & 37.50\% \\
MP4 & 35.29\% & 90.17\% & 60.00\% & 80.00\% \\
PDF & 98.80\% & 95.71\% & 60.00\% & 86.66\% \\
    \bottomrule
\end{tabular}
}
\end{table}

\section{Library Influence}
\label{sec:LibraryInfluence}
To evaluate the impact of different libraries on the quality of
generator, we conduct experiments across four target formats. Specifically, we
use GPT-4 to construct generators, specifying the library to be used in the
prompt, such as ``\textit{You must use cv2 to create this Python
generator.}'' For each format, we select two libraries capable of generating
files in the corresponding format.

The results are presented in \T~\ref{tab:diffLibrary}. In most cases, different
libraries exhibit large variations in feature coverage, as observed in the cases
of JPG, MP4, and PDF. Notably, combining multiple libraries leads to higher
overall feature coverage because their complementary functionalities enable the
construction of more sophisticated generators. 

\begin{table}[!htbp]
    \small
    \footnotesize
    \centering
    \caption{Feature coverage achieved by using different libraries.}
    \vspace{-5pt}
    \label{tab:diffLibrary}
  \setlength{\tabcolsep}{3pt}
    \resizebox{0.7\linewidth}{!}{
    \begin{tabular}{c |c | c | c }
    \toprule
    Format & Library & Feature Cov & ValidNum/InvalidNum \\ 
    \midrule
\multirow{3}{*}{JPG} & PIL         & 252                  & 40/0                            \\
& cv2         & 337                  & 40/0                            \\
& Unlimited   & \textbf{984}                  & 130/0                           \\
\midrule
\multirow{3}{*}{MP4} & cv2         & \textbf{471}                  & 23/11                           \\
& moviepy     & -                    & -                               \\
& Unlimited   & 290                  & 19/15                           \\
\midrule
\multirow{3}{*}{PDF} & fpdf        & 353                  & 27/0                            \\
& PyPDF2      & 72                   & 18/4                            \\
& Unlimited   & \textbf{559}                  & 50/1                            \\
\midrule
\multirow{3}{*}{TIFF} & PIL        & 164                  & 27/0                            \\
& tifffile   & 161                  & 22/2                            \\
& Unlimited  & \textbf{387}                  & 48/8    \\
    
    \bottomrule
\end{tabular}
}
\end{table} 

\begin{figure*}[t] 
    \centering
    \begin{minipage}[t]{0.2\textwidth}
        \centering
        \includegraphics[width=\textwidth]{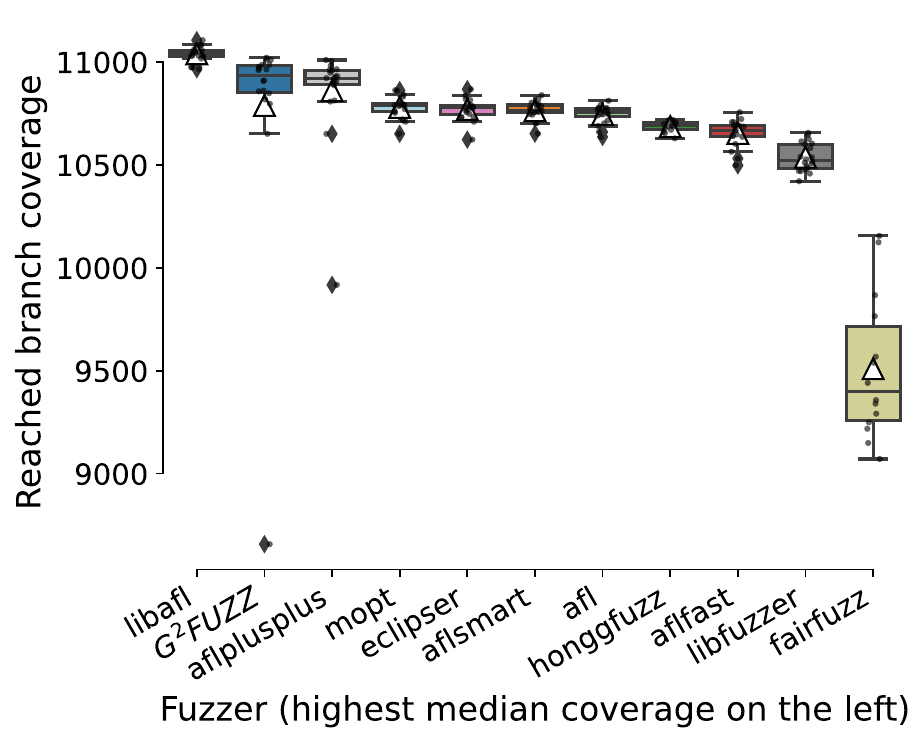}
        \vspace{-17pt}
        \subcaption{harfbuzz}
    \end{minipage}
    \begin{minipage}[t]{0.2\textwidth}
        \centering
        \includegraphics[width=\textwidth]{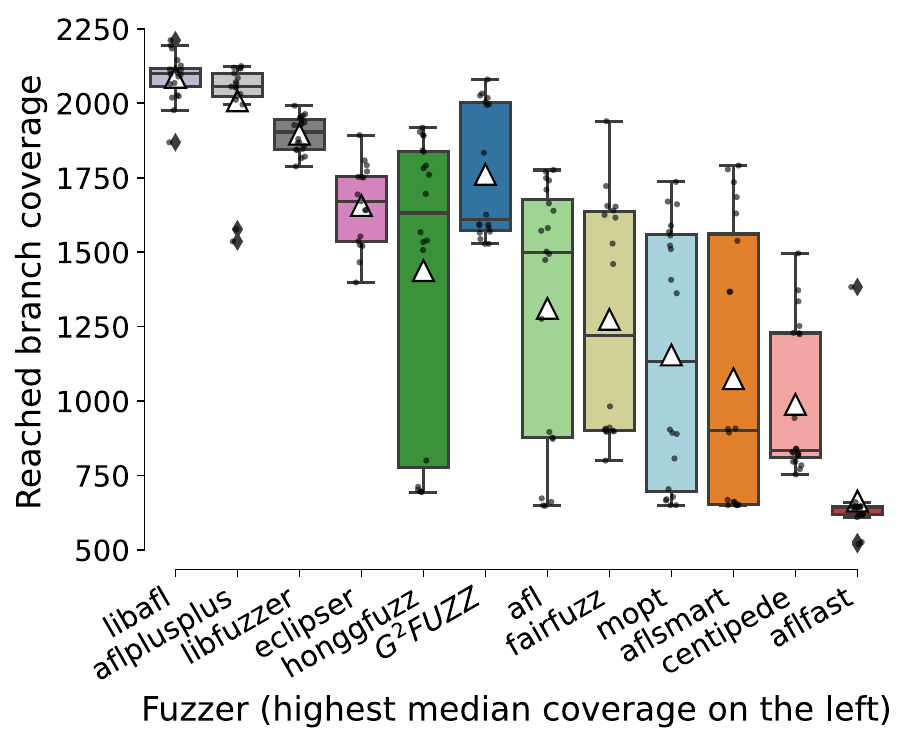}
        \vspace{-17pt}
        \subcaption{lcms}
    \end{minipage}
    \begin{minipage}[t]{0.2\textwidth}
        \centering
        \includegraphics[width=\textwidth]{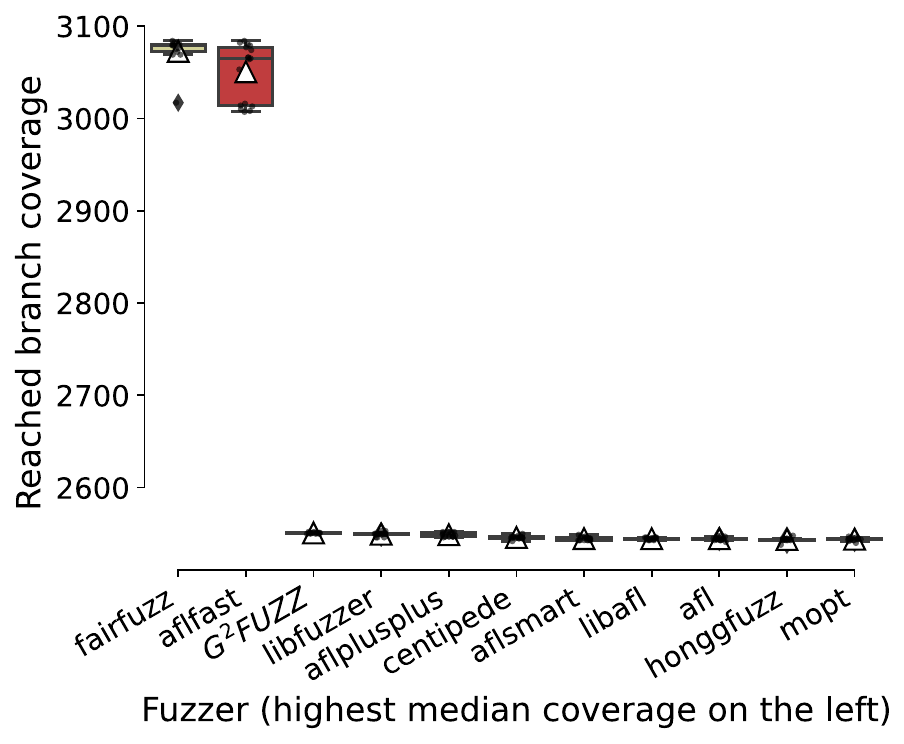}
        \vspace{-17pt}
        \subcaption{libjpeg}
    \end{minipage}
    \begin{minipage}[t]{0.2\textwidth}
        \centering
        \includegraphics[width=\textwidth]{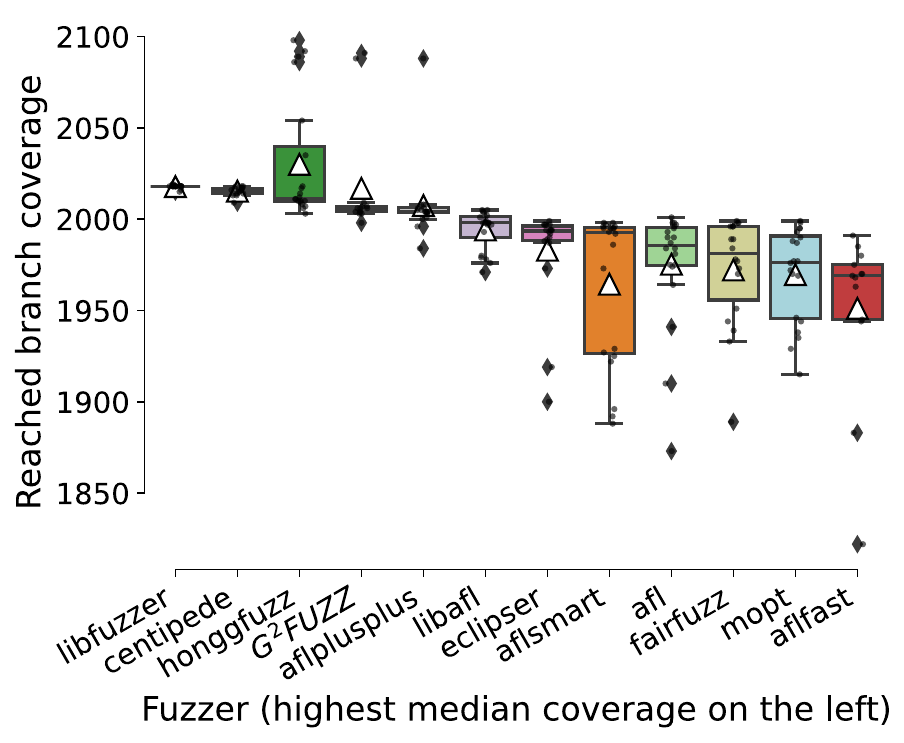}
        \vspace{-17pt}
        \subcaption{libpng}
    \end{minipage}
    \begin{minipage}[t]{0.2\textwidth}
        \centering
        \includegraphics[width=\textwidth]{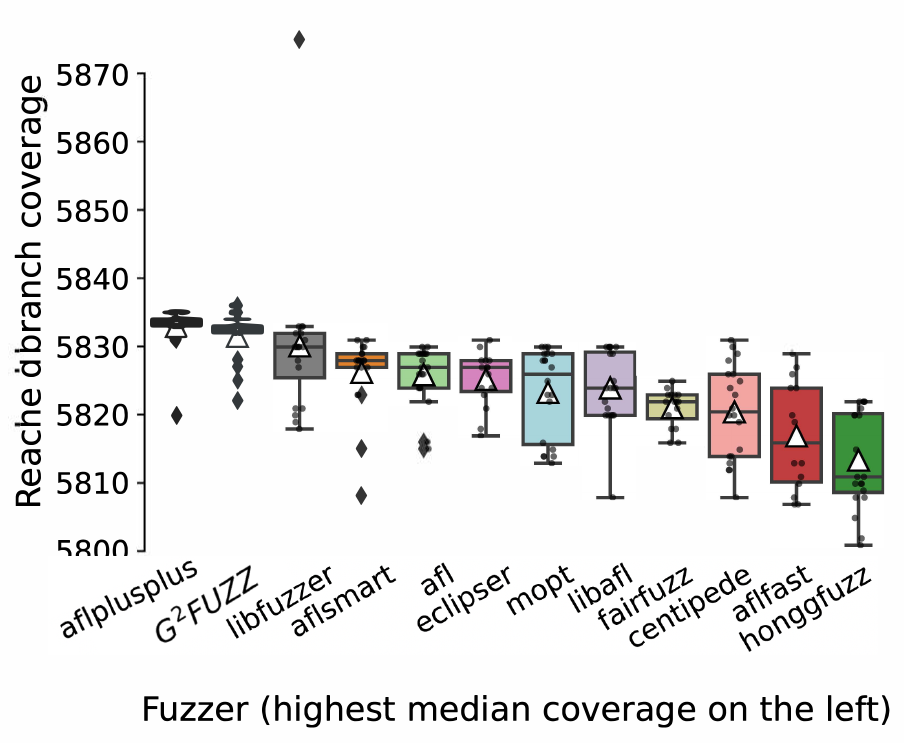}
        \vspace{-17pt}
        \subcaption{openssl}
        \vspace{-10pt}
    \end{minipage}
    \begin{minipage}[t]{0.2\textwidth}
        \centering
        \includegraphics[width=\textwidth]{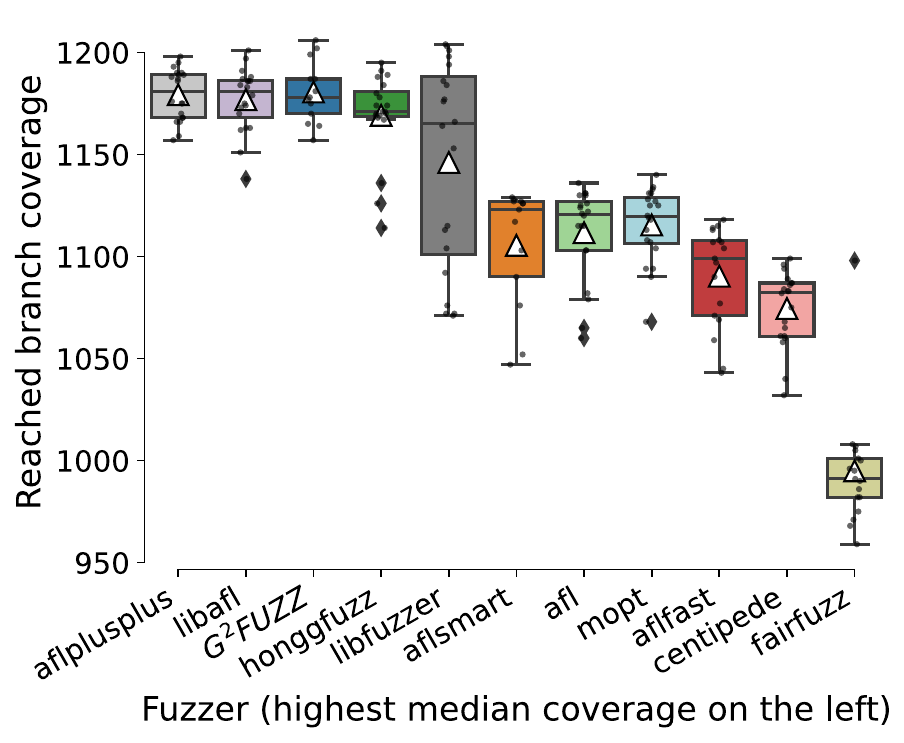}
        \vspace{-17pt}
        \subcaption{woff2}
        \vspace{-10pt}
    \end{minipage}
    \begin{minipage}[t]{0.2\textwidth}
        \centering
        \includegraphics[width=\textwidth]{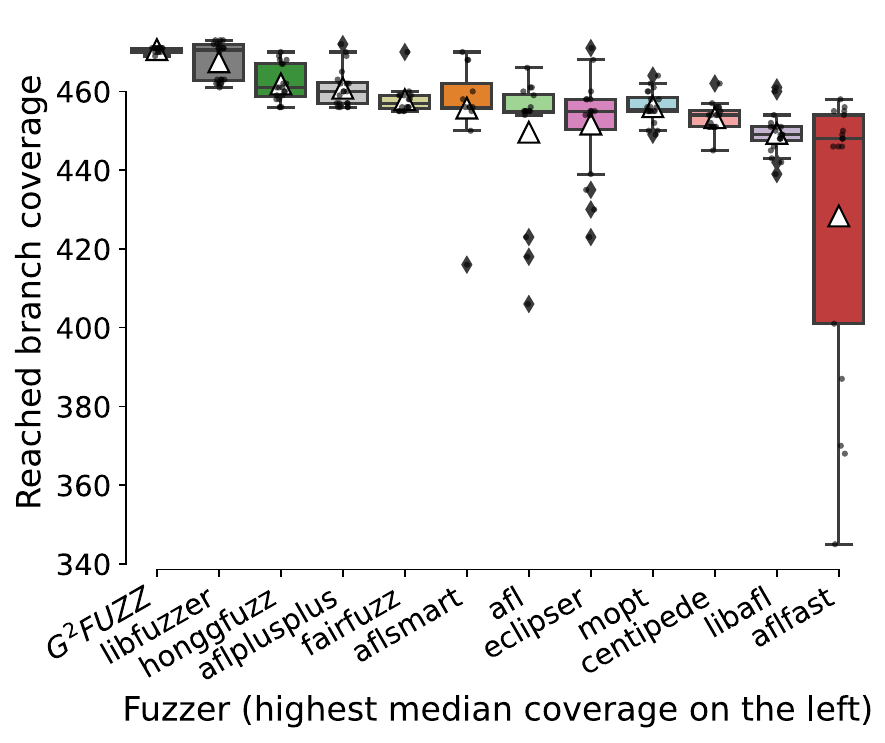}
        \vspace{-17pt}
        \subcaption{zlib}
        \vspace{-5pt}
    \end{minipage}
    \vspace{-5pt}
    \caption{Code coverage distributions achieved in a FuzzBench experiment.}
    \vspace{-20pt}
    \label{fig:fuzzbenchOthers}
\end{figure*}

\begin{figure}[!htbp]
    \centering
    \vspace{-5pt}
    \includegraphics[width=0.9\linewidth]{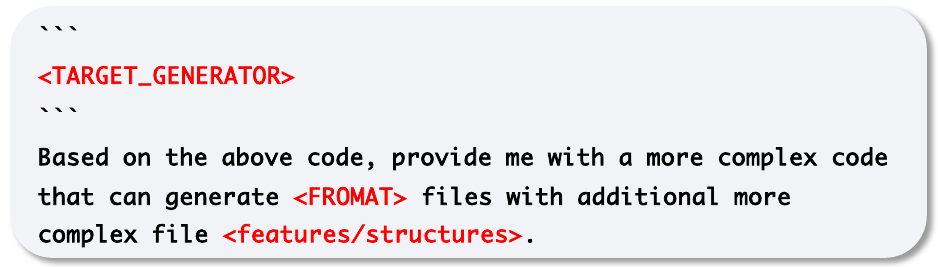}
    \caption{The prompt for random mutation.}
    \vspace{-10pt}
    \label{fig:randomMutation}
\end{figure}

\begin{table}[!htbp]
    \centering
    \caption{Benchmark programs selected from UNIFUZZ, FuzzBench, and MAGMA.}
    \label{tab:allPrograms}
    \vspace{-5pt}
    \resizebox{0.9\linewidth}{!}{
  \setlength{\tabcolsep}{3pt}
    \begin{tabular}{llll}
        \toprule
        \textbf{UNIFUZZ} & \textbf{FuzzBench} & \textbf{MAGMA} \\
        \midrule
        gdk-pixbuf-pixdata & bloaty\_fuzz\_target & libpng\_read\_fuzzer \\
        jhead & freetype2-2017 & read\_rgba\_fuzzer \\
        mp3gain & harfbuzz-1.3.2 & tiffcp \\
        ffmpeg & lcms-2017-03-21 & pdf\_fuzzer \\
        tiffsplit & libjpeg-turbo-07-2017 & pdfimages \\
        pdftotext & libpcap\_fuzz\_both & pdftoppm \\
        mp42aac & libpng-1.2.56 & sndfile\_fuzzer \\
        flvmeta & openssl\_x509 &  \\
        imginfo & vorbis-2017-12-11 &  \\
        exiv2 & woff2-2016-05-06 &  \\
         & zlib\_zlib\_uncompress\_fuzzer &  \\
        \bottomrule
    \end{tabular}
    }
\end{table}

\begin{figure}[h!] 
    \vspace{-20pt}
    \centering
    \includegraphics[width=0.4\textwidth]{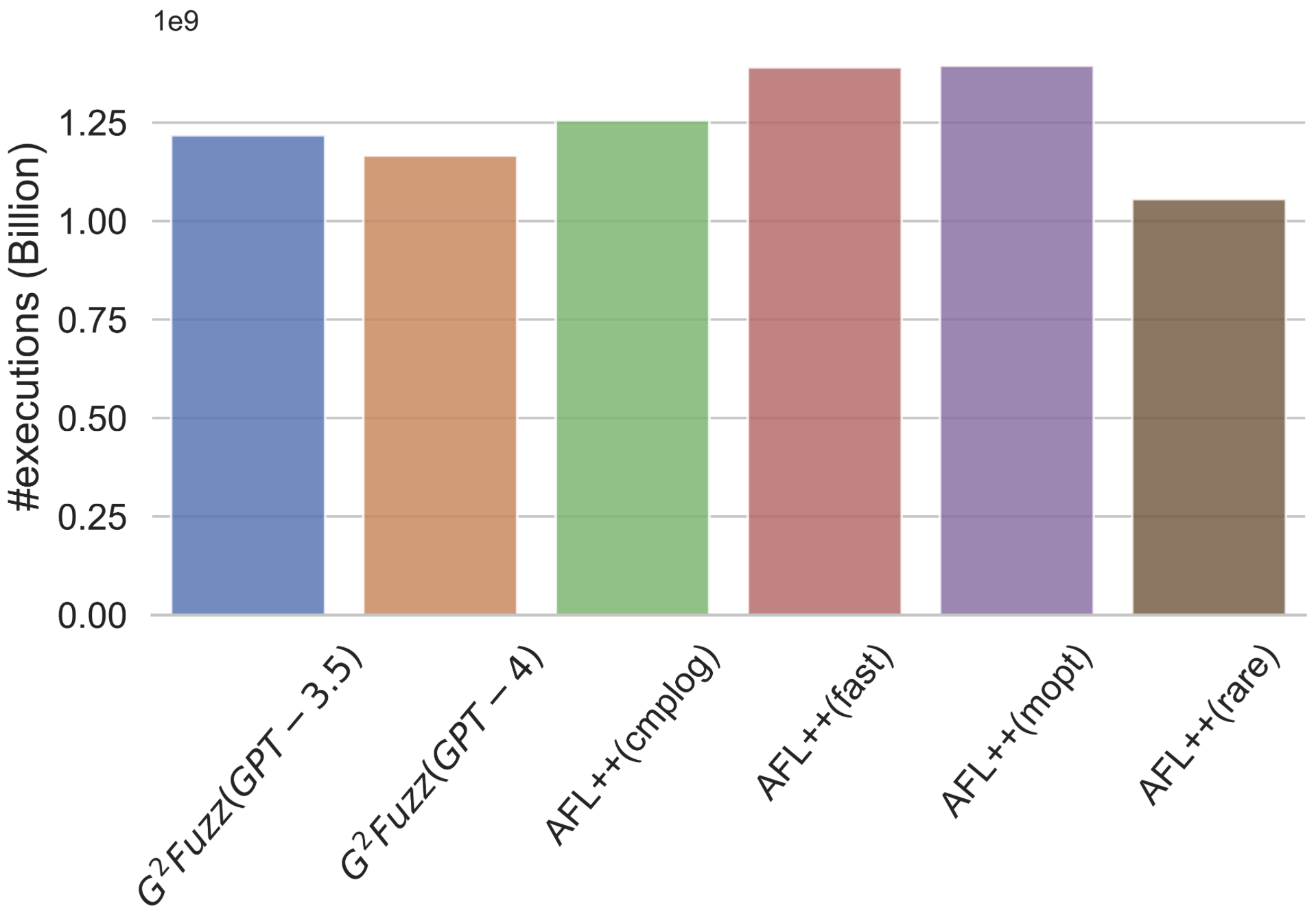}
    \caption{The total throughput of different fuzzers running for 24 hours.}
    \vspace{-20pt}
    \label{fig:throughput}
\end{figure}

\begin{figure}[!h] 
    \centering
    \includegraphics[width=0.49\textwidth]{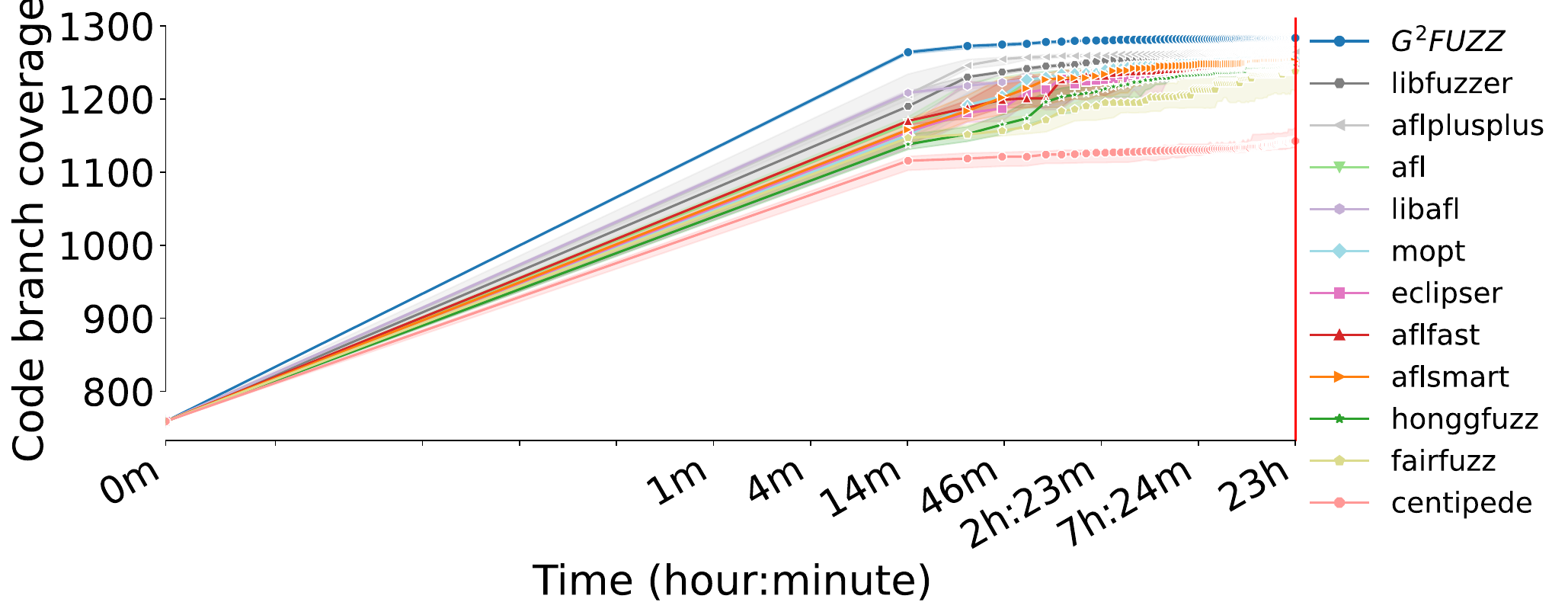}
    \vspace{-10pt}
    \caption{Average Code Coverage Evolution Over Time for \textit{vorbis_decode_fuzzer}.}
    \vspace{-5pt}
    \label{fig:fuzzbenchVorbis}
\end{figure}

\begin{table}[!h]
    \small
    \footnotesize
    \centering
    \caption{The additional analysis time of LLMGenFuzz. The unit is seconds.}
    \label{tab:motivation}
    \resizebox{0.85\linewidth}{!}{
    \begin{threeparttable}
    \begin{tabular}{c | c | c | c | c}
    \toprule
    Programs & Group I & Group II & Group III & Average(ExtraPCT\tnote{1}\ )\\
    \midrule
    lcms &  131 &  299 &  158 &  196(0.22\%)\\
    woff2 &  431 &  434 &  220 &  361(0.42\%)\\
    vorbis &  552 &  329 &  194 &  358(0.41\%)\\
    freetype2 &  817 &  788 &  410 &  671(0.77\%)\\
    libpcap &  261 &  173 &  220 &  218(0.25\%)\\
    bloaty &  743 &  990 &  1,043 &  925(1.06\%)\\
    harfbuzz &  465 &  651 &  282 &  466(0.54\%)\\
    libjpeg-turbo &  575 &  163 &  118 &  285(0.33\%)\\
    libpng  &  379 &  158 &  47 &  194(0.22\%)\\
    openssl &  1,259 &  189 &  551 &  666(0.77\%)\\
    zlib &  220 &  161 &  110 &  163(0.19\%)\\
    \bottomrule
    \end{tabular}
    \begin{tablenotes}
    \item[1] \textbf{ExtraPCT}: The percentage of the additional analysis time compared to the 23 hours of fuzzing time.
    
    \end{tablenotes}
    \end{threeparttable}
    }
    \vspace{-10pt}
    \label{tab:extraTime}
\end{table} 

\begin{table}[!h]
    \small
    \footnotesize
    \centering
    \caption{The median code coverage achieved by \tool at 23 hours and 23 hours and 45 mins.}
    \label{tab:motivation}
    \resizebox{0.85\linewidth}{!}{
    \begin{tabular}{c | c | c | c }
    \toprule
    \multirow{2}{*}{Programs}
     & \multirow{2}{*}{23 hours} & 23 hours and & \multirow{2}{*}{Diff}\\
     && 45 minutes &\\
    \midrule
    bloaty_fuzz_target & 6,377.0  &  6,377.0 &  0 \\
    freetype2_ftfuzzer &  11,630.0 &  11,630.0 &  0 \\
    harfbuzz_hb-shape-fuzzer & 10,935.5  & 10,935.5  & 0  \\
    lcms_cms_transform_fuzzer   & 1,610.0  &  1,610.0 & 0 \\
    libjpeg-turbo_libjpeg_turbo_fuzzer & 2,551.0  & 2,551.0  &  0 \\
    libpcap_fuzz_both &  3,003.0 &  3,003.0 &  0 \\
    libpng_libpng_read_fuzzer &  2,006.0 &  2,006.0 &  0 \\
    openssl_x509 &  5,833.0 &  5,833.0 &  0 \\
    vorbis_decode_fuzzer  &  1,283.5 &  1,283.5 &  0 \\
    woff2_convert_woff2ttf_fuzzer &  1,178.0 &  1,178.0 & 0  \\
    zlib_zlib_uncompress_fuzzer &  471.0 &  471.0 & 0  \\
    \bottomrule
    \end{tabular}
    }
    \vspace{-5pt}
    \label{tab:23hVS22h45min}
\end{table}


\end{document}